\documentclass[reprint,prb,twocolumn,amsmath,amssymb]{revtex4-1}

\usepackage{mathrsfs}
\usepackage{mathtools}
\usepackage{xfrac}
\usepackage{color}

\usepackage{graphicx}
\usepackage{dcolumn}
\usepackage{bm}

\newcommand{\anni}[2]{\hat{\psi}_{#1}(#2)}
\newcommand{\crea}[2]{\hat{\psi}^{\dagger}_{#1}(#2)}

\newcommand{\anniB}[1]{\hat{\Psi}(#1)}

\newcommand{\ket}[1]{ | #1 \rangle }

\newcommand{\order}[1]{\langle #1 \rangle }

\newcommand{\vet}[1]{\vec{#1}}

\usepackage{hyperref}

\begin{document}


\title{Symmetry analysis of translational symmetry broken 
  density waves: \\ application to hexagonal lattices in two dimensions}

\author{J. W. F. Venderbos}
\affiliation{%
Department of Physics, Massachusetts Institute of Technology, Cambridge, Massachusetts 02139, USA
}%

\date{\today}

\begin{abstract}
In this work we introduce a symmetry classification for electronic density waves which break translational symmetry due to commensurate wave vector modulations. The symmetry classification builds on the concept of extended point groups: symmetry groups which contain, in addition to the lattice point group, translations that do not map the enlarged unit cell of the density wave to itself, and become ``non-symmorphic''-like elements. Multi-dimensional representations of the extended point group are associated with degenerate wave vectors. 
Electronic properties such as (nodal) band degeneracies and topological character can be straightforwardly addressed, and often follow directly. To further flesh out the idea of symmetry, the classification is constructed so as to manifestly distinguish time-reversal invariant charge (i.e., site and bond) order, and time-reversal breaking flux order. For the purpose of this work we particularize to spin-rotation invariant density waves. 
As a first example of the application of the classification we consider the density waves of a simple single- and two-orbital square lattice model. The main objective, however, is to apply the classification to two-dimensional (2D) hexagonal lattices, specifically the triangular and the honeycomb lattices. The multi-component density waves corresponding to the commensurate $M$-point ordering vectors are worked out in detail. To show that our results generally apply to 2D hexagonal lattices, we develop a general low-energy SU(3) theory of (spinless) saddle-point electrons. 
\end{abstract}

\pacs{pacs}

\maketitle


\section{Introduction}

Interactions between electrons in solids are responsible for a large variety of symmetry broken electronic phases. Unconventional superconductivity and anti-ferromagnetism are two canonical examples. From the perspective of weak coupling, the anti-ferromagnet is an example of a density wave: a spatial modulation of electronic spin density characterized by a finite (generally commensurate) propagation vector, breaking translational symmetry. In general, apart from spin density, other observables involving particle-hole pairs such as charge density, orbital density, or current density, can acquire finite wave-vector modulations and break translational symmetry, giving rise to many possible unconventional and exotic density wave states.   

A systematic way to study unconventional density waves is to classify them according to their symmetry properties. Density wave states are condensates of particle-hole pairs and can be classified by specifying the angular momentum and spin of the particle-hole pair~\cite{nayak00,platt13}, in close analogy to unconventional superconductivity~\cite{sigrist91}. For instance, the familiar charge and spin density waves are both $s$-wave condensates. Distinct angular momentum channels are labeled by representations of the symmetry group of the crystal. In case of particle-hole condensates, contrary to a superconductor, spin and angular momentum are not tied together by Fermi statistics.

Knowledge of the symmetry of density wave states is essential for understanding the properties of materials, in particular for connecting theory to experiment. In addition, from a modern condensed matter perspective, determining the symmetry of an electronic phase is particularly relevant, as symmetries can give rise to topological electronic states~\cite{hasan10,qi11,ando15,bernevig,franz}. The presence of a symmetry, such as time-reversal and/or parity symmetry, can protect or prohibit a topological phase, which may be either gapped or gapless. A number of works have specifically explored the role of lattice symmetries~\cite{fu11,hsieh12,fang12,slager12,alexandradinata14}. In the context of density wave states, a number of proposals for realizing interaction-induced topological particle-hole condensates have been made in recent years~\cite{raghu08a,yang10,castro11,hsu11,li15}, which have set the stage for the rapidly growing and evolving field of interacting topological phases. Motivated by the exciting possibility of electrons condensing into exotic collective phases, possibly characterized by large nonzero angular momentum and spontaneously generated charge- or spin-currents, in this paper we develop a symmetry analysis of density waves with finite commensurate wave-vector modulation. 

When particle-hole condensation occurs at finite wave vector, breaking translational symmetry, the crystal symmetry group is reduced to the group of the wave vector. The symmetry classification proposed in this paper naturally takes translational symmetry breaking into account. Instead of a reduced symmetry group, it is defined in terms of a symmetry group called the \emph{extended} point group. The essential feature of the extended point group can be summarized by noting that it treats a given set of ordering wave vectors on similar footing with angular momentum channels. As a result, extended point groups provide a means to classify translational symmetry broken density waves, in much the same way as ordinary point groups provide a means to classify different angular momentum channels~\cite{sigrist91,nayak00,platt13}. In particular, multidimensional representations signal degeneracies in both cases. This property is particularly useful in case of inequivalent but symmetry-related ordering wave vectors. Another way to think of extended point groups is to consider them as the point groups of a mathematically enlarged unit cell, i.e., a unit cell which supports the physically enlarged unit cell associated with condensation at given wave vector (or set of wave vectors). The extended point group contains translations that do not map the enlarged unit cell to itself. These are the translations broken by density wave formation. 

In the next section we start by introducing the symmetry classification based on the notion of extended point groups. The classification of particle-hole condensates manifestly distinguishes two types of order: time-reversal invariant charge order and time-reversal breaking flux order. The former class consists of diagonal site order of the form $\Delta  \sim \order{\hat{\psi}^\dagger_i \hat{\psi}_i}$, and off-diagonal bond order of the form $\Delta \sim \text{Re}\,\order{\hat{\psi}^\dagger_i \hat{\psi}_j}$, where $i,j$ label lattice sites. The second class, flux order, corresponds to dynamically generated orbital currents, i.e., order of the form $ \Delta  \sim \text{Im}\,\order{\hat{\psi}^\dagger_i \hat{\psi}_j}$. Importantly, instead of the imaginary part of bond expectation values, the symmetry classification is defined in terms of fluxes: the sum of phases around a closed lattice plaquette. This preserves gauge invariance by construction. In this work, we only consider spin-singlet or spin-rotation invariant density wave order.

As a simple first example, we revisit the density wave states of the square lattice by an application of the symmetry classification. As a second example, we apply the symmetry classification to a two-orbital square lattice model, which can be viewed as a basic description of certain iron-pnictide materials~\cite{haule08,raghu08b}. Given these useful examples, our main goal is to apply to apply the symmetry classification to lattices with hexagonal symmetry, in particular the triangular and honeycomb lattices. Hexagonal lattices have two sets of special commensurate wave vectors: the $K$  and $M$ points. The emphasis will be on the latter due their special commensurability and threefold degeneracy. Degeneracies can give rise to multi-component orders. We find two sets of $M$-point density waves common to lattices with hexagonal symmetry: a set of conventional $s$ waves (i.e., charge density waves) and a set of time-reversal odd $d$ waves (charge-current density waves). The mean-field ground state of the rotationally symmetric triple-$M$ $d$-wave state is a Chern insulator. 

As a next step, using extended lattice symmetries, we study the hexagonal $M$-point density waves focusing only on relevant low-energy electronic degrees of freedom. This will allow us to cast the results demonstrated for the triangular and honeycomb lattices in a more general form. Due to the threefold degeneracy of the $M$-points, the low-energy electrons come in three flavors. We obtain and discuss the low-energy SU(3) theory governing these three-flavor electrons. In particular, it is straightforward to demonstrate that $s$- and $d$-density waves correspond to nesting instabilities. We address the quasiparticle gap structures of these density waves and show that the triple-$M$ $d$-wave state is always associated with nonzero Chern number.

We summarize the organization of this paper as follows. In Sec.~\ref{sec:class}, the symmetry classification is defined and applied to the square lattice (Secs.~\ref{ssec:sq} and~\ref{ssec:2orbsq}). In Sec.~\ref{sec:condens}, the symmetry classification is applied to the hexagonal lattices, including a further characterization of selected density wave states. In Sec.~\ref{sec:lowenergy}, the analysis is complemented by focusing on the relevant low-energy degrees of freedom. We summarize and conclude in Sec.~\ref{sec:summary}. A number of appendixes collect details of results presented in the main text.


\section{Symmetry classification of condensates\label{sec:class}}

Condensation of particle-hole pairs at finite commensurate wave vector implies the breaking of translational symmetry and an enlargement of the crystal unit cell. To develop a classification that takes this feature into account in a systematic way, we describe the system in terms of a mathematically enlarged unit cell, chosen so as to support the physically enlarged unit cell of the density waves we want to study. The choice of the mathematically enlarged unit cell is determined by the ordering wave vectors relevant to the particular physical system. All possible patterns of translational symmetry breaking can then be described within the new enlarged unit cell. The symmetry group of the (mathematically) enlarged unit cell is the \emph{extended} point group, which consists of all point group elements of the Bravais lattice supplemented with the translations that do not map the enlarged unit cell onto itself, i.e., the translations that can be broken in the density wave state. As a result, these are an extension of the ordinary point groups with additional composite elements. Naturally, the extended point group depends on the set of ordering vectors.

As an example, consider ordering at wave vector $\vet{Q}= (\pi,\pi) $ of the square lattice. This breaks the translation $T(\vet{a}_1)$ over the unit vector $\vet{a}_1$ and doubles the unit cell. Clearly, the translation $T(\vet{a}_1)$ does not map the doubled unit cell onto itself, but instead connects the two sites of the enlarged cell. The translations $T(\vet{a}_1+\vet{a}_2)$ and $T(\vet{a}_1-\vet{a}_2)$ do map the doubled unit cell to itself and generate the group of unbroken (or invariant) translations. Adding $T(\vet{a}_1)$ to the square point group $C_{4v}$ gives the extended group $C'_{4v}$, where one prime is meant to indicate that one translation has been added. The group $C'_{4v}$ is the symmetry group of the enlarged unit cell, and is treated as any other ordinary point group. In particular, its character table follows from its algebraic structure. Note that in the extended group $C'_{4v}$ the translation $T(\vet{a}_1)$ is its own inverse, since $2T(\vet{a}_1)$ is an invariant translation. Similarly, $T(\vet{a}_2)$ is equivalent to $T(\vet{a}_1)$ as it can be written as $T(\vet{a}_1)-T(\vet{a}_1-\vet{a}_2)$. In Appendix~\ref{app:gt}, we collect some point group essentials, specifically in relation to extended point groups, and list character tables of symmetry groups. 

In our classification we distinguish three types of orderings: site (or charge) order, bond order, and flux order. Flux order originates from imaginary bond order amplitudes breaking time-reversal symmetry, but since we only aim to distinguish gauge inequivalent orders, we define time-reversal breaking orders by fluxes.  We adopt a real-space lattice approach to classify all types of ordering according to lattice symmetry. The set of ordering vectors, specified a priori, determines the size of the (mathematically) enlarged unit cell. In this work we will consider cases where the unit cell is at most quadrupled. The enlarged unit cell contains $n_s$ sites $\{ s_i \}^{n_s}_{i=1} $, $n_b$ bonds $\{ b_i \}^{n_b}_{i=1} $, and $n_\phi$ fluxes $\{ \phi_i \}^{n_\phi}_{i=1} $. The fluxes $\phi_i$ are associated with each plaquette of the lattice. The extended point group operations $g$ permute the elements of the vectors $\vet{s}$, $\vet{b}$, and $\vet{\phi}$. Writing the permutation matrix as $P^{s}(g)$, and similarly for bonds and fluxes, we have (repeated indices are summed), 
\begin{gather}
s'_{i} = P^{s}_{ij}(g) s_j, \quad b'_{i} = P^{b}_{ij}(g) b_j, \quad \phi'_{i} = P^{\phi}_{ij}(g) \phi_j.
\end{gather} 
It is important to bear in mind that fluxes change sign under reflections, and therefore elements of $P^{\phi}(g)$ acquire a minus sign when $g$ is a reflection. In each of the three cases the set of all permutations defines a representation of the extended point group, which we call $\mathcal{P}_{s}$ for site order and similarly for bond and flux order. 

The representations $\mathcal{P}_{s}$, $\mathcal{P}_{b}$, and $\mathcal{P}_{\phi}$ are reducible and can be decomposed into a sum of irreducible representations. We take this decomposition to define the symmetry classification of all particle-hole condensates of a given type of order. Specifically, the particle-hole condensates are basis functions of the irreducible representations. The transformation properties of the condensates under lattice symmetries directly follow from the symmetry of the representation. In particular, the dimensionality of the representation is equal to the number of symmetry-related partner density waves. Therefore, the construction in terms of symmetry representations is useful to study degeneracies and multiple-$Q$ ordering scenarios. 

Even though the classification itself is defined in terms of a real-space construction, ultimately we are interested in explicit momentum-space expressions of particle-hole condensates, in order to, for instance, study mean-field Hamiltonians. The symmetry classification serves this purpose by providing an exhaustive list of all condensates supported by the ordering vectors, which are obtained using a simple and straightforward construction. The symmetry properties of the condensates, automatically delivered by the classification, can be used to derive the condensate functions systematically, even though in simple cases they follow directly. In addition, the distinction between time-reversal even and gauge invariant time-reversal odd orderings is naturally formulated in the real space construction. A further benefit of the present classification is that the symmetry of density waves, labeled by representations of the symmetry group, can be directly used to derive phenomenological Landau theories. Phenomenological models can be derived and analyzed simply based on the symmetry of the order parameter, and do not require knowledge of condensate functions. In particular, multi-component Landau theories typically give rise to distinct composite or subsidiary orders, which can be directly obtained using the symmetry classification. This will be demonstrated below in the context of a simple example (Sec.~\ref{ssec:2orbsq}).  

The extended point group structure is hierarchical in the sense that the bare point group (without translations) is a proper subgroup. Therefore, all representations of the extended point group can be decomposed into representations of the bare point group. This decomposition contains information regarding the spectral effects of the density waves, as we will see in the following. Furthermore, this decomposition can be used to define and study multiple-$Q$ ordering in case the extended point group representation describes inequivalent wave vector components. 

We now apply the symmetry classification to a single- and two-orbital square lattice model. The next section is then devoted to the triangular and honeycomb lattices. 

\subsection{Application I: Simple square lattice\label{ssec:sq}}

\begin{figure}
\includegraphics[width=0.8\columnwidth]{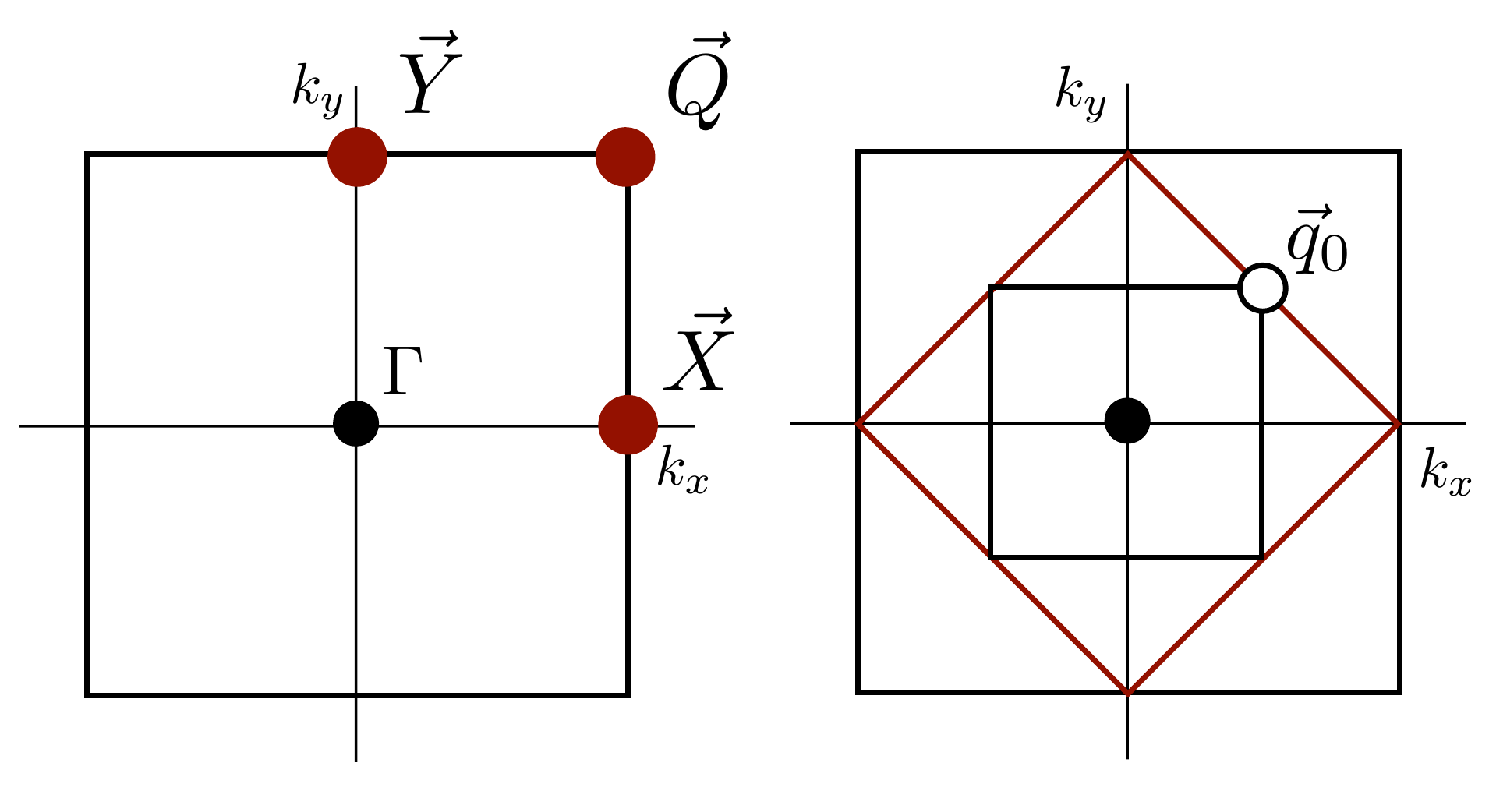}
\caption{\label{fig:bz_square} (Left) Brillouin zone (BZ) of the square lattice. The momenta $\vet{Q} = \pi(1,1)/a$, $\vet{X} = \pi(1,0)/a$, and $\vet{Y} = \pi(0,1)/a$ are marked by bold red dots. (Right) Red rotated square inscribed in the square lattice BZ marks the nested Fermi surface at half-filling. Inner black square represents the reduced Brillouin zone for multiple-$Q$ ordering and $\vet{q}_0 = (\pi,\pi)/2a$ denotes the location of the degeneracy point of the $d_{x^2-y^2}$ density-wave state.}
\end{figure}

The starting point is fixing the ordering vectors. We consider the set of ordering vectors given by the three momenta 
\begin{gather}
\vet{Q}=\frac{\pi}{a}(1,1), \; \vet{X}=\frac{\pi}{a}(1,0),  \;  \vet{Y}=\frac{\pi}{a}(0,1),
\end{gather}
which are shown in Fig.~\ref{fig:bz_square}. Each of these ordering vectors is half of a reciprocal lattice vector, e.g. $2\vet{Q} = 0$, which is an expression of their commensurability. 
The unit cell is quadrupled and as a consequence the three translations $T(\vet{a}_1)$, $T(\vet{a}_2)$ and $T(\vet{a}_3)\equiv T(\vet{a}_1+\vet{a}_2)$ become members of the extended square symmetry group $C'''_{4v}$. The character table of $C'''_{4v}$ is reproduced in Appendix~\ref{app:gt}~\cite{serbyn13} .

The enlarged unit cell of the square lattice has $n_s=n_\phi = 4$ and $n_b=8$. Using the recipe of the symmetry classification, we construct the permutation representations $\mathcal{P}_{s}$, $\mathcal{P}_{b}$, and $\mathcal{P}_{\phi}$ and decompose into irreducible representations. For site order we find the following decomposition 
\begin{gather} \label{eq:sqsite}
\mathcal{P}_{s} = A_1 + B'_2 + E_5.
\end{gather}
The first term is the trivial representation. The second and third term are representations specific to $C'''_{4v}$ and correspond to translational symmetry breaking. From the symmetry of the $B'_2$ state we see that it describes site order at wave vector $\vet{Q}$, i.e., a staggered charge density wave. The representation $E_5$ is two-dimensional and describes a doublet of striped charge orderings with wave vectors $\vet{X}$ and $\vet{Y}$.

In case of bond order, we find that $\mathcal{P}_{b}$ is decomposed as
\begin{gather}  \label{eq:sqbond}
\mathcal{P}_{b} = A_1 + B_1 + E'_1 + E_3 + E_5.
\end{gather}
The square lattice unit cell (as opposed to the enlarged one) contains two bonds and the first to terms $A_1 + B_1  $ describe the translationally invariant bond modulations. The remaining representations correspond to translational symmetry breaking and are all two-dimensional. The $E'_1$ doublet is a set of staggered $p$-waves with ordering vector $\vet{Q}$. The $E_5$ doublet is symmetry equivalent to the site ordered doublet of stripes with ordering vectors $\vet{X}$ and $\vet{Y}$.

For the flux order representation $\mathcal{P}_{\phi}$ we find the decomposition 
\begin{gather} \label{eq:sqflux}
\mathcal{P}_{\phi} = A_2 + A'_2 + E_2.
\end{gather}
The fluxes necessarily break time-reversal and reflection symmetries, which explains the absence of a fully invariant term $A_1$. The representation $A'_2$ corresponds to a state of staggered fluxes with ordering vector $\vet{Q}$ and $d$-wave ($d_{x^2-y^2}$) structure. This staggered flux or $d$-density wave has a long history in the context of spin liquid~\cite{affleck88,hsu91} and cuprate pseudogap physics~\cite{chakravarty01}. 
The doublet $E_2$ corresponds to staggered flux stripes with wave vectors $\vet{X}$ and $\vet{Y}$.

The bond ordered states of Eq.~\eqref{eq:sqbond} derive from bonds connecting nearest neighbor sites. We can apply the same method to the diagonal bonds, connecting next-nearest neighbors. Including the diagonal bonds in the bond vector $\vet{b}$ yields the decomposition
\begin{gather*}  \label{eq:checkbond}
\mathcal{P}_{b} = 2 A_1 + B_1 + B_2 + A'_1 + B'_2 + E'_1 + E_2 + 2E_3 + E_5.
\end{gather*}
The state with $A'_1$ symmetry has $d$-wave ($d_{xy}$) structure and ordering vector $\vet{Q}$. It is the time-reversal even $d$-wave partner of the flux ordered state with $A'_2$ symmetry. 

The notable density wave states of this example are summarized in Table~\ref{tab:square}. To conclude this example of square lattice density waves, we demonstrate how symmetry can be used to identify topologically nontrivial states. In two dimensions and disregarding spin the most common nontrivial state is the Chern insulating state. Time-reversal symmetry and reflection symmetry each force the Chern number to be zero~\cite{fang12} and we are therefore limited to flux ordered states. The $A'_2$ state is odd under all reflections (see Appendix~\ref{app:gt} for $C'''_{4v}$ character table) but it is even under reflections combined with a translation $T(\vet{a}_1)$, which is sufficient to enforce zero Chern number. Equivalently, time-reversal symmetry is preserved up to a translation. Similarly, one cannot form a flux ordered state out of the two $E_2$ partners which manifestly breaks all reflections. As a result, a Chern insulating state cannot be formed within a single density wave channel.  

We can alternatively consider mixed representation states. Inspection of the characters of the representations $A'_1$ ($d_{xy}$) and $A'_2$ ($d_{x^2-y^2}$) leads to the conclusion that a combination of these breaks time-reversal symmetry and all reflections. Hence, a $d+id$ state can have nonzero Chern number. Both $d$-waves have semi-metallic gapless spectra, a combination of the two of the form $d+id$ has a gapped mean-field spectrum. One can start from the (imaginary) $d_{x^2-y^2}$ state and show that admixture of the (real) $d_{xy}$ state gaps out the nodal degeneracies, or vice versa. Indeed the resulting insulating state has nonzero Chern number and is associated with a spontaneous quantum Hall (QH) effect~\cite{laughlin98,kotetes08}. In addition, it was shown that when the spin degree of freedom is included, the spin-dependent superposition $d+ i\sigma d$ describes a quantum spin Hall phase~\cite{kane05,hsu11}.

\begin{table}[t]
\centering
\begin{ruledtabular}
\begin{tabular}{cccl}
Representation & Angular momentum  &  $Q$-vector  & $\Theta$ \\ 
\hline
$A'_1$  & $d_{xy}$ & $\vet{Q}$ & $+$ \\
$A'_2$  & $d_{x^2-y^2}$ & $\vet{Q}$ &  $-$\\
$B'_2$  & $s$-wave & $\vet{Q}$& $+$\\
$E'_1$  & $p$-wave & $\vet{Q}$ &  $+ $\\
$E_2$  & $p$-wave & $\vet{X}$,$\vet{Y}$ & $-$ \\
$E_3$  &$p$-wave & $\vet{X}$,$\vet{Y}$ & $+$
\end{tabular}
\end{ruledtabular}
 \caption{Summary of the square lattice symmetry classification of particle-hole condensates with (at most) quadrupled unit cell. Condensates are labeled by the representation of the extended point group. Their angular momentum, ordering vector, and transformation under time reversal $\Theta$ are listed. (Note that the origin is chosen at the center of a square plaquette.)}
\label{tab:square}
\end{table}

\subsection{Application II: Two-orbital square lattice\label{ssec:2orbsq}}

As a second example, we consider a modified square lattice model with an internal orbital degree of freedom. To be concrete, we take a square lattice with $\ket{xz}$ and $\ket{yz}$ orbitals at each site. This may be considered as the simplest two-orbital model capturing the essential features of iron-pnictide materials~\cite{haule08,raghu08b}. 

We focus on ordering at the same ordering vectors as in the previous example. Due to the two orbitals per site, the number of sites is effectively doubled. In constructing the permutation representation $\mathcal{P}_{s}$ the transformation of the orbital degree of freedom has to be taken into account. The two orbitals $\{\ket{xz},\ket{yz} \}$ are odd under twofold rotation, however, since we are interested in \emph{density} wave order, the sign change should not be accounted for. What is important is whether a symmetry exchanges the orbitals. The fourfold rotation and the diagonal reflection exchange the two orbitals. Keeping this in mind we construct $\mathcal{P}_{s}$ and find
\begin{gather} \label{eq:2orbsite}
\mathcal{P}_{s} = A_1 + B_1+B'_2 + A'_2+2E_5.
\end{gather}
Before we proceed we should note that this result is basis dependent. Basis dependence is a consequence of including the orbital degree of freedom. We have chosen the basis $\{\ket{xz},\ket{yz} \}$, which diagonalizes the vertical reflections, but we could have chosen the basis $\ket{\pm} = \ket{xz}\pm \ket{yz}$, which diagonalizes the diagonal reflections. Alternatively, we can choose the basis $\ket{\pm i} = \ket{xz}\pm i\ket{yz}$, which diagonalizes the fourfold rotations. Defining the Pauli matrix $\tau^3 = \pm 1$ to describe the orbital degree of freedom, then the different basis choices correspond to the condensate expectation value $\order{\hat{\psi}^\dagger_a \hat{\psi}_b} \sim \tau^i_{ab}$. In the basis $\ket{\pm} = \ket{xz}\pm \ket{yz}$ (corresponding to $\tau^1$), for instance, the site order decomposition reads as
\begin{gather} \label{eq:2orbsite2}
\mathcal{P}'_{s} = A_1 + B_2+B'_2 + A'_1+E_4+E_5.
\end{gather}
The representations $E_{4,5}$ both describe two-component charge density modulations with wave vector $\vet{X}$ and $\vet{Y}$. In Eqs.~\eqref{eq:2orbsite} and~\eqref{eq:2orbsite2}, one of the two-component representations also has orbital density modulations (given by $\tau^3$ and $\tau^1$, respectively). In the spin channel (i.e., multiplying by $\sigma^3$) these two-component representations would correspond to the spin density waves observed in the iron-pnictide materials~\cite{stewart11}. The representation $B'_2$ is easily seen to describe a charge density wave with wave vector $\vet{Q}$ (see Table~\ref{tab:square}), and corresponds to a composite order associated with a collinear biaxial spin density wave~\cite{fernandes15}. Within the present approach, this may be seen from taking the product $E_5 \times E_5 = A_1 + B_1 + A'_2 + B'_2$, similarly for $E_4$. 

The bond order representation depends on which bonds are included. For simplicity we consider only nearest-neighbor bonds, which for the $\{\ket{xz},\ket{yz} \}$ orbitals implies that only intra-orbital bond connections are possible. The permutation representation is then constructed and decomposed as
\begin{gather} \label{eq:2orbbond}
\mathcal{P}_{b} = 2A_1 +2B_1+ 2E'_1 + 2E_3+2E_5.
\end{gather}
It is worth mentioning the $B_1$ term, which does not correspond to translational symmetry breaking, but instead breaks the fourfold rotations. As  a result, it corresponds to the composite nematic order associated with magnetic fluctuations~\cite{si08,fang09,xu08,qi08,fernandes10,fernandes12}. That nematic order is accompanied with orbital order may be seen from~\eqref{eq:2orbsite}: the $B_1$ term corresponds to orbital order.

Since we restricted this simple example to intra-orbital bonds, each square plaquette has two fluxes. The flux order decomposition reads as
\begin{gather} \label{eq:2orbflux}
\mathcal{P}_{\phi} = A_2 +B_2+ A'_2 + B'_2+2E_2.
\end{gather}
We single out the term with $A'_2$ symmetry, which is a $d$-density wave at wave vector $\vet{Q}$ (see Table~\ref{tab:square}). Comparing symmetries we find that it corresponds to the composite order parameter associated with a spin-vortex crystal state~\cite{fernandes15}. Showing this explicitly within the present framework would require a more careful consideration of spin, which is beyond the scope of this work. 


\section{Multi-component density waves of hexagonal lattices \label{sec:condens} }

We now come to the main aim of this paper: application of the symmetry classification to the hexagonal triangular and honeycomb lattices. In the first part of this section, we apply the symmetry classification introduced in the previous section to both cases. We will be primarily concerned with the three-fold degenerate $M$-point ordering vectors and we only give a brief description of honeycomb lattice $K$-point density waves. The honeycomb lattice $K$ points are special because the Dirac nodes are located at these momenta~\cite{castroneto09}. The Dirac nodes are protected by the symmetries of the $K$-point little group $C_{3v}$, and can only be gapped by lowering the symmetry, or by coupling the two nodes (i.e., breaking translational symmetry). In the second part, we take a closer look at the properties of the density wave states with $M$-point modulation that correspond to nesting and Pomeranchuk instabilities at special filling.

\subsection{Symmetry classification applied to hexagonal lattices\label{ssec:hexagonal}}

\begin{figure}
\includegraphics[width=\columnwidth]{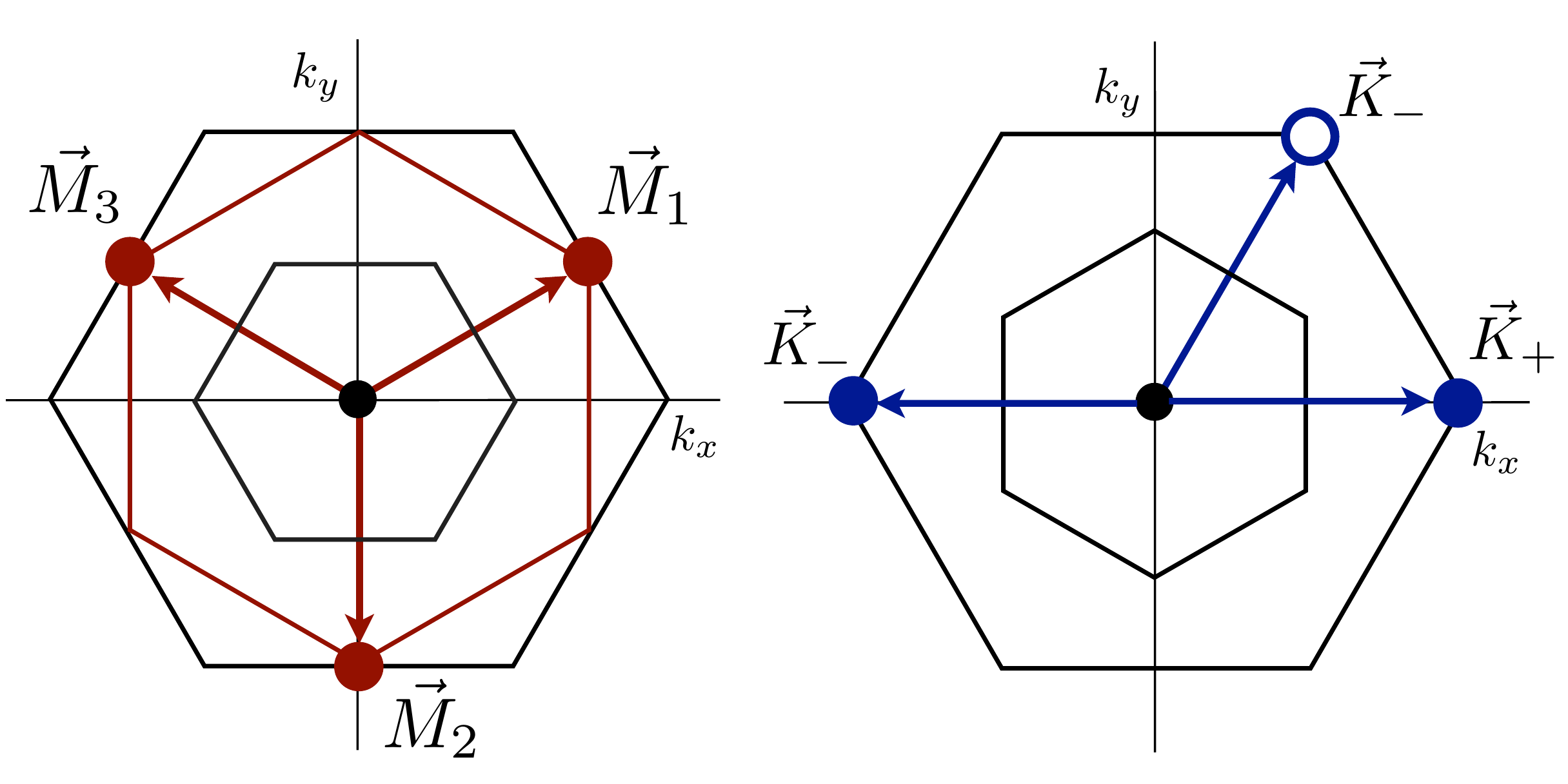}
\caption{\label{fig:bz_hexa} (Left) Large black hexagon is the Brillouin zone (BZ) of hexagonal lattices. Bold red dots denote the $M$-points and the red hexagon connecting these $M$-points is the Fermi surface at van Hove filling. Small black hexagon denotes the folded BZ corresponding to unit cell quadrupling. (Right) Hexagonal BZ with blue dots denoting the $K$-points, and the small (rotated) black hexagon is the reduced BZ corresponding to ordering at $\vet{K}_\pm$.}
\end{figure}

The BZ of hexagonal lattices, shown in Fig.~\ref{fig:bz_hexa}, has two sets of special momenta (apart from the zone center $\Gamma$). These are \emph{(i)} the corners of the hexagon (i.e., the $K$-points), and \emph{(ii)} the centers of the edges (i.e., the $M$-points). All of these wave vectors give rise to commensurate orderings. 

The first set of ordering vectors consists of the inequivalent $M$-point vectors, which we label $\vet{M}_\mu$, $\mu=1,2,3$. They are shown in Fig.~\ref{fig:bz_hexa} (left) and given by
\begin{gather} \label{eq:trimpoints}
\vet{M}_{1,3}  = \frac{\pi}{a\sqrt{3}} ( \pm \sqrt{3} , 1 ) , \quad \vet{M}_2  = \frac{2\pi}{a\sqrt{3}} ( 0, -1 ).
\end{gather}
The $M$-point momenta are half of a reciprocal lattice vector ($2\vet{M}_\mu=0$), and are related by $\pm \vet{M}_1 \pm \vet{M}_2 \pm \vet{M}_3 =0$. To describe ordering at the $M$-point wave vectors we need to consider a quadrupled the unit cell. Three translations are broken and added to the point group, yielding the extended group $C'''_{6v}$~\cite{hermele08}. 

The triangular and the honeycomb lattice band structures have the property that the Fermi surface is perfectly nested for specific filling fractions, assuming only nearest neighbor hopping. For the triangular lattice the special filling is $n=3/4$, whereas for the honeycomb lattice it is $n=1/2\pm 1/8$. The nesting vectors are $\vet{M}_\mu$, which connect the van Hove singularities located at $\vet{M}_\mu$ since $\vet{M}_1 = \vet{M}_2-\vet{M}_3$ (similarly for other combinations). Therefore, condensation at these wave vectors is directly related to nesting instabilities of the triangular and honeycomb lattices. 

The second set, the $K$-points, consists of two inequivalent momenta $\vet{K}_+$ and $\vet{K_-}$, which are shown in Fig.~\ref{fig:bz_hexa}(right). These ordering vectors satisfy $2\vet{K}_+=  \vet{K}_-$ and $3\vet{K}_+=0$, which implies $\vet{K}_- = -\vet{K}_+$. In this paper we only consider $K$-point condensation for the honeycomb lattice, since the honeycomb lattice Dirac nodes are located at the $K$-points. 
Ordering at the $K$-point vectors leads to a tripled unit cell and two broken translations which are added to the point group. The extended point group of hexagonal lattice $K$-point ordering is $C''_{6v}$~\cite{basko08}. 

\subsubsection{Triangular lattice}

\begin{table}[t]
\centering
\begin{ruledtabular}
\begin{tabular}{cccl}
Representation & Angular momentum  &  $Q$-vector  & $\Theta$ \\ 
\hline
$E_2$  & $d$-wave & $\vet{Q}=0$ & $+$ \\
$F_1$  & $s$-wave & $\vet{M}_1$, $\vet{M}_2$, $\vet{M}_3$ &  $+$\\
$F_2$  & $d$-wave & $\vet{M}_1$, $\vet{M}_2$, $\vet{M}_3$  &  $-$\\
\end{tabular}
\end{ruledtabular}
 \caption{List of common hexagonal lattice $s$- and $d$-wave density waves at wave vectors $Q=0$ and $Q=M$. The density waves summarized in this table exist in all hexagonal lattices, the $F_1$ and $F_2$ representations correspond to the van Hove nesting instabilities in each case. }
\label{tab:morder}
\end{table}

We first take the simplest case of the triangular lattice and derive all possible $M$-point orderings. Due to the quadrupling of the unit cell we have $n_s=4$, $n_b=12$ and $n_\phi=8$. The representations of the extended group $C'''_{6v}$ which reflect the breaking of translational symmetry are all three dimensional. This is consistent with the threefold degeneracy of the ordering vectors $\vet{M}_\mu$. For site ordering we find the decomposition
\begin{gather} \label{eq:trisitem}
\mathcal{P}_{s} = A_1 + F_1.
\end{gather}
Apart from the trivial $A_1$ state we find three charge density waves associated with an $M$-point vector $\vet{M}_\mu$ which are components of the representation $F_1$. Since charge density waves are $s$-wave states, we can associate the $s$-wave character to $F_1$ representation. 

For triangular lattice bond order we find the decomposition
\begin{gather} \label{eq:tribondm}
\mathcal{P}_{b} = A_1 + E_2  + F_1  + F_3 + F_4,
\end{gather}
which, in addition to the trivial state, contains a $d$-wave doublet given by $E_2$. These $d$ waves preserve translations and inversion but break rotational symmetry. The two-fold degeneracy of the $d$-wave channel is a consequence of hexagonal symmetry. Nematic $Q=0$ order can arise via a Pomeranchuk instability~\cite{pomeranchuk59,valenzuela08}. We find bond density waves with $F_1$ symmetry and therefore refer to them as extended $s$-wave states. The representations $F_3$ and $F_4$ are both odd under inversion and therefore correspond to states of $p$-wave type. 

Finally, the flux order decomposition is given by
\begin{gather} \label{eq:trifluxm}
\mathcal{P}_{\phi} =  A_2  + B_1  +  F_2 +  F_3. 
\end{gather}
The state with $B_1$ symmetry is an $f$-wave flux ordered state which preserves translational invariance. The representation $F_2$ describes three translational symmetry broken charge-current $d$-density waves. These $d$-waves can be viewed as analogs of the square lattice $d_{x^2-y^2}$ state. This will become clear below, then we study them in more detail (see also Fig.~\ref{fig:trif2}). Of particular interest will be the case of triple-$M$ $d$-wave ordering: simultaneous ordering of the three $F_2$ components. By invoking symmetry we can already predict that triple-$M$ order has the proper symmetry to support a spontaneous Quantum Hall effect. The representation $F_2$ can be decomposed into irreducible representations of the subgroup $C_{6v}$. We find $F_2=A_2+ E_2$ and a chiral state with $A_2$ symmetry is compatible with nonzero Chern invariant.

\subsubsection{Honeycomb lattice}

\begin{table}[t]
\centering
\begin{ruledtabular}
\begin{tabular}{ccccc}
Representation & Angular momentum & $Q$-vector  & Sublattice & $\Theta$ \\ 
\hline
$B_2$  & $s$-wave & $\vet{Q}=0$ & $\tau^3$ & $+$ \\
$A_2$  & $f$-wave & $\vet{Q}=0$ & $\tau^3$  &$-$\\
$E'_1$  & & $\vet{K}_+$, $\vet{K}_-$ & $\tau^1$,$\tau^2$ & $+$\\
$E'_2$  & & $\vet{K}_+$, $\vet{K}_-$ & $\tau^1$,$\tau^2$ &  $- $\\
$G'$   & & $\vet{K}_+$, $\vet{K}_-$ & $\tau^0$,$\tau^3$  & $+$ \\
\end{tabular}
\end{ruledtabular}
 \caption{List of density waves specific to the honeycomb lattice at wave vectors $\vet{Q}=0$ and $\vet{Q}=\vet{K}_\pm$. Note that the $f$-wave state originates from next-nearest neighbor bond order waves. Here the $\tau^i$ are Pauli matrices acting on the honeycomb sublattice degree of freedom. Note that, for instance, $\tau^3$ has $B_2$ symmetry. }
\label{tab:korder}
\end{table}

We start by deriving the $M$-point ordered states. The honeycomb lattice has two triangular sublattices, the $A$ and $B$ sublattices, and in case of $M$-point ordering we have $n_s = 8$, $n_b = 12$ and $n_\phi = 4$. For site order we have the decomposition
\begin{gather}  \label{eq:hexasitem}
\mathcal{P}^{M}_{s} = A_1 + B_2 + F_1 + F_4.
\end{gather}
The appearance of the $B_2$ state is due to the sublattice structure and corresponds to inversion symmetry breaking sublattice polarized charge order. The $s$-wave states with $F_1$ symmetry are the honeycomb lattice equivalent of the charge density wave states of the triangular lattice with same symmetry. 

The bond order decomposition is the same as for the triangular lattice and we thus have
\begin{gather}  \label{eq:hexabondm}
\mathcal{P}^{M}_{b} = A_1 + E_2 + F_1 + F_3 + F_4 . 
\end{gather}
Since both lattices have the same symmetry, the nature of the states in the decomposition is the same. In particular, the $E_2$ doublet corresponds to honeycomb lattice nematic $d$-waves and $F_1$ is an $s$-wave triplet. 

Honeycomb lattice flux order is decomposed into irreducible representations as 
\begin{gather} \label{eq:hexafluxm}
\mathcal{P}^{M}_{\phi} = A_2  + F_2.
\end{gather}
The only translational symmetry broken ordering is in the $d$-wave channel with $F_2$ symmetry. These orders are the honeycomb lattice version of the $d$-wave triangular lattice orders with $F_2$ symmetry. Therefore, our argument based on symmetry, that a nontrivial ground state with nonzero Chern number is in principle allowed, applies to these orders in the same way. 

One might wonder whether the Haldane state, the arrangement of fluxes that average to zero over the unit cell~\cite{haldane88}, is present in the decomposition. The answer is no, since the Haldane state involves next-nearest neighbor bonds. We can, however, obtain the Haldane state by simply using the results of the triangular lattice, given that the honeycomb lattice is composed of two triangular sublattices. The Haldane state is translationally invariant, and the only translationally invariant triangular flux ordered state has $B_1$ or $f$-wave symmetry. Multiplying it with the staggered representation $B_2$ coming from \eqref{eq:hexasitem}, we obtain a state with $A_2$ symmetry, which is listed as $f$-wave state in Table~\ref{tab:korder}. It is this staggered $f$-wave state that corresponds to the Haldane state. Hence, the this example of the Haldane state shows nicely how the symmetry classification applied to the triangular lattice may be ``nested'' to obtain density waves of the honeycomb lattice (connecting next-nearest neighbors).

We proceed to apply the symmetry classification to density wave order at $K$ points. The site, bond and flux order representations are decomposed with respect to irreducible representations of $C''_{6v}$. For site order we find
\begin{gather} \label{eq:hexasiteK}
\mathcal{P}^{K}_{s} = A_1 + B_2 + G',
\end{gather}
which only differs with respect to~\eqref{eq:hexasitem} in the translational symmetry broken part $G'$, as it should. The representation $G'$ is fourfold degenerate and thus describes a set of four site ordered charge density waves modulated by $\vet{K}_\pm$. In a low-energy description of the honeycomb lattice Dirac node electrons, these charge density waves affect the Dirac nodes by moving them in momentum space, implying that they couple to the low-energy electrons as gauge fields~\cite{gopalakrishnan12}. This can be inferred directly from the structure of $G'$ by decomposing it into irreducible representations of $C_{6v}$, giving $G'=E_1+ E_2$. Both these representations are two dimensional, implying that they correspond to a two-component gauge field $\sim \vet{A} = (A_x,A_y)$. Therefore, we directly find the low-energy spectral effect of these density waves from the symmetry classification.

Bond ordering at the $K$ points is given by the orderings
\begin{gather}  \label{eq:hexabondK}
\mathcal{P}^{K}_{b} = A_1 + E_2 + E'_1 + G'.
\end{gather}
In addition to the representation $G'$, the decomposition contains the doublet $E'_1$. It describes the two K\'ekule-type bond modulation patterns of the honeycomb lattice. Honeycomb lattice distortions of this type were found to give a complex mass (i.e., two real degrees of freedom) to the low-energy Dirac electrons~\cite{hou07}. The real and imaginary parts of the complex K\'ekule mass correspond to the two $E'_1$ states. This can be established in a straightforward way by deriving the extended point group symmetry transformation properties of Dirac fermion bilinears~\cite{basko08,juan13}. 

To conclude, we consider flux order with $K$-point modulations. The flux order decomposition reads as
\begin{gather} \label{eq:hexafluxK}
\mathcal{P}^{K}_{\phi} = A_2 + E'_2 .
\end{gather}
The flux order term $E'_2$ can be considered the time-reversal breaking version of $E'_1$ K\'ekule bond order and describes modulated flux patterns~\cite{castro11}. Symmetry dictates that these modulated flux ordered states cannot correspond to Dirac masses, as there are no mass bilinears with corresponding symmetry. Instead, in the low-energy theory both $E'_2$ states lead to a valley splitting, energetically shifting the Dirac nodes (i.e., valleys). This inequivalence of time-reversal related partners is consistent with time-reversal symmetry breaking implied by flux order. Decomposing the $E'_2$ representation in terms of irreducible $C_{6v}$ representations we find $A_2+B_2$. Both states are metallic at charge neutrality. Based on simple symmetry arguments, we know that the $A_2$ state allows for nonzero Hall conductivity~\cite{sun08} and we therefore identify it as a topological Fermi liquid, reproducing the result of Ref.~\onlinecite{castro11}. (Note that the structure of the $E'_2$ representation is important: the two basis states labeled $A_2$ and $B_2$ cannot map to each other under translations.)

The results for $K$-point order are summarized in
Table~\ref{tab:korder}. By applying the symmetry classification to honeycomb lattice electronic
order with $K$-point modulations, we have demonstrated how it provides a direct identification and simple characterization of these orders.


\subsection{Density waves of hexagonal lattices \label{ssec:condtri}}

Next, we take a more detailed look at some of the hexagonal density waves by writing down explicit expressions for the particle-hole condensates. We select the nematic $d$-waves with $E_2$ symmetry as well as the three-component charge density and flux density waves with $F_1$ and $F_2$ symmetry. The motivation for selecting these states is that they correspond to generic nesting and Pomeranchuk instabilities. This is demonstrated and explained in the next section.

In case of hexagonal symmetry, the $d$ waves are degenerate partners of the $E_2$ representation. As a result, a $Q=0$ $d$-wave condensate is in general a superposition of the two $d$-wave form factors. The appropriate triangular lattice $L=2$ angular momentum functions, denoted $\lambda_{d_1}(\vet{k})$ and $\lambda_{d_2}(\vet{k})$, are given in Table~\ref{tab:trifunctions} of Appendix~\ref{app:gt}, and a general $d$-wave condensate takes the form
\begin{gather} 
\order{\crea{\sigma }{\vet{k}} \anni{\sigma' }{\vet{k}}} = [ \Delta_{1}\lambda_{d_1}(\vet{k}) + \Delta_{2}\lambda_{d_2}(\vet{k})] \delta_{\sigma\sigma'}.  \label{eq:e2tri}
\end{gather}
Hermiticity requires the two independent order parameters $\Delta_1$ and $\Delta_2$ to be real. Spin-singlet $d$-wave order gives rise to a deformation of the Fermi surface, breaking rotational symmetry but preserving inversion symmetry. The two components $(\Delta_1,\Delta_2)$ therefore constitute a nematic director. Such nematic Fermi surface deformations originating from a Pomeranchuk instability have been studied in the context of doped graphene~\cite{valenzuela08}.

Now, let us proceed to the density waves with finite $M$-point ordering vectors. We examine the charge density waves with $F_1$ symmetry and the charge-current (flux) density waves with $F_2$ in more detail. First, we look at the triangular lattice and then consider the honeycomb lattice.

\begin{figure}
\includegraphics[width=0.9\columnwidth]{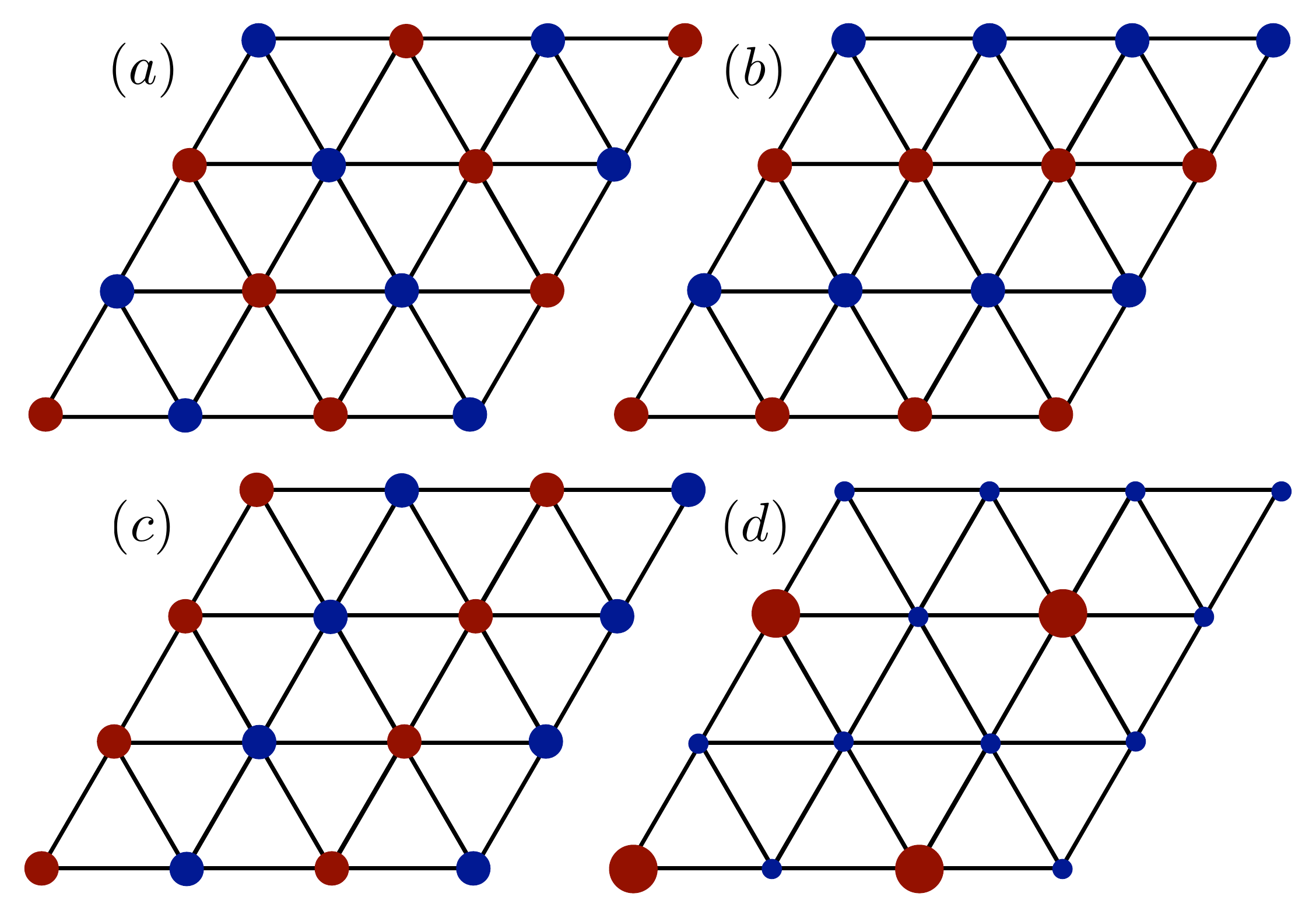}
\caption{\label{fig:trif1} Real space picture of the site order with $F_1$ symmetry. (a)--(c) Show the components with $\vet{M}_1$, $\vet{M}_2$, and $\vet{M}_3$ modulations, respectively. (d) Shows equal amplitude triple-$M$ order with $A_1$ symmetry. }
\end{figure}

There are two kinds of density waves with $F_1$ symmetry: site ($s$-wave) and bond (extended $s$-wave) order. An expression for the charge density $s$ waves is directly obtained by giving each component $\vet{M}_\mu$ ($\mu=1,2,3$) momentum-independent strength, expressed as
\begin{gather} \label{eq:f1tri}
\order{\crea{\sigma}{\vet{k}+\vet{M}_\mu }\anni{\sigma' }{\vet{k}}} = \Delta_\mu\;  \delta_{\sigma\sigma'}. 
\end{gather}
The order parameters $\Delta_\mu$ must be real. Triple-$M$ order is realized when all components have equal amplitude, $\Delta_1=\Delta_2=\Delta_3$. Site order with $F_1$ symmetry is shown in Fig.~\ref{fig:trif1} where (a)--(c) are the three components of $F_1$ and Fig.~\ref{fig:trif1}(d) shows triple-$M$ order. Triple-$M$ order breaks translational symmetry but preserves all elements of the group $C_{6v}$. This follows from decomposing $F_1$ into representations of $C_{6v}$, which gives $F_1=A_1+E_2$. One can take the $A_1$ term to define triple-$M$ order. 

\begin{figure}
\includegraphics[width=0.9\columnwidth]{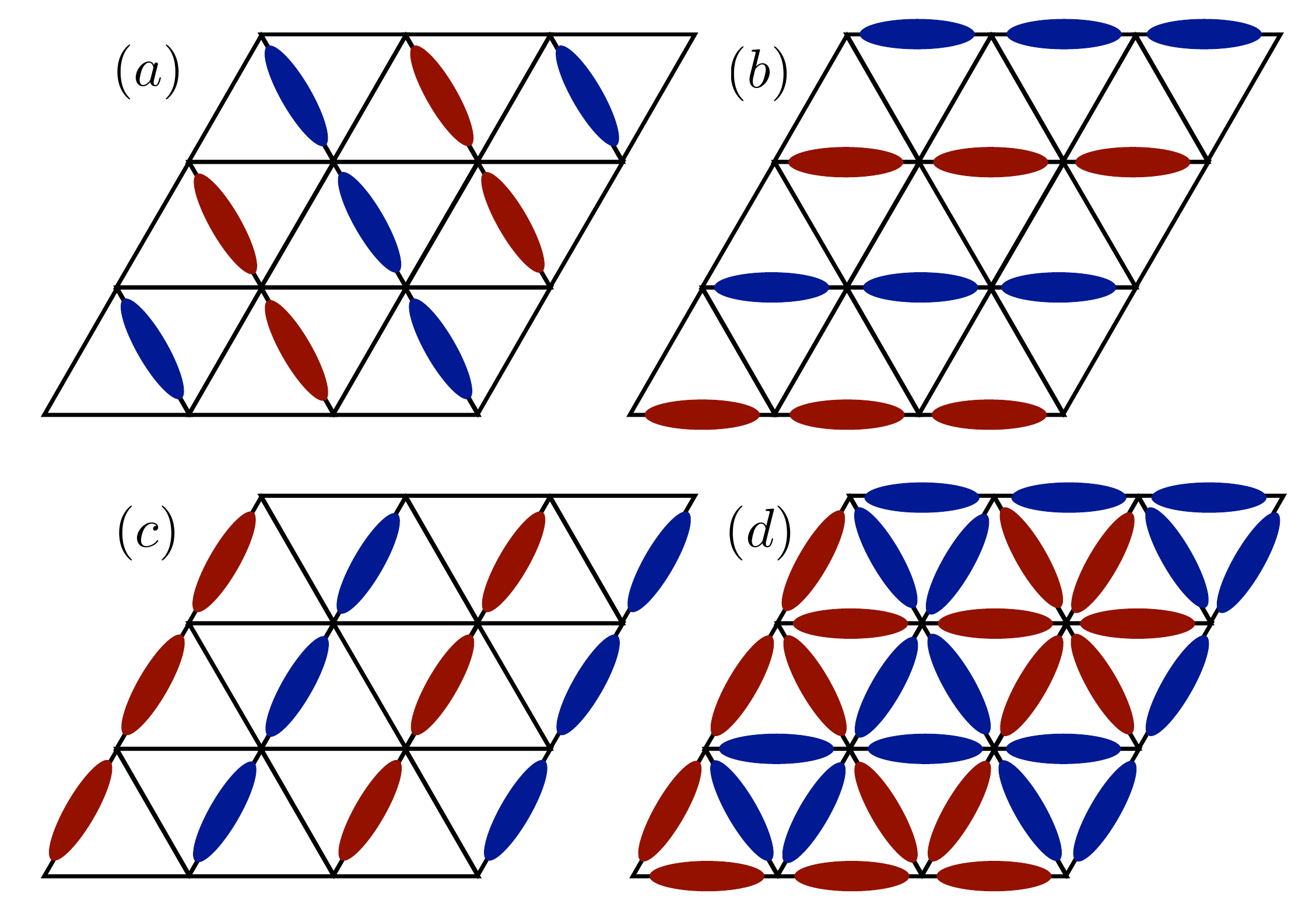}
\caption{\label{fig:trif1bond} Real space picture of the bond order with $F_1$ symmetry. Again, (a)--(c) show the individual components as in Fig.~\ref{fig:trif1} and (d) shows equal amplitude triple-$M$ order.}
\end{figure}

To obtain the condensate functions of $F_1$ bond order we consider momentum-dependent form factors transforming as representations of the little group of each $\vet{M}_\mu$. Taking $\vet{M}_1$ as an example, the (extended) $s'$-wave functions are given by $\cos k_2$ and $\cos k_1+\cos k_3$ (where $k_i = \vet{k}\cdot \vet{a}_i$). Only $\cos k_2$ is consistent with time-reversal symmetry according to arguments outlined in Appendix~\ref{ssec:compat} [specifically Eq.~\eqref{eq:compat}]. A similar analysis applies to $\vet{M}_2$ and $\vet{M}_3$, and in terms of the condensate functions $\Delta_\mu(\vet{k})$ defined by $\order{\crea{\sigma}{\vet{k}+\vet{M}_\mu }\anni{\sigma' }{\vet{k}}}  = \Delta_\mu(\vet{k}) \delta_{\sigma\sigma'}$ we find
\begin{gather} \label{eq:f1bondtri}
\Delta_1(\vet{k})  =  \Delta_{1}  \cos k_2  , \nonumber \\
\Delta_2(\vet{k})  =   \Delta_{2} \cos k_3 , \nonumber \\
\Delta_3(\vet{k})  =   \Delta_{3}  \cos k_1 .
\end{gather}
The three components $\Delta_\mu$ are shown in Figs.~\ref{fig:trif1bond}(a)--(c) and uniform triple-$M$ order is shown in Fig.~\ref{fig:trif1bond}(d).

\begin{figure}
\includegraphics[width=0.9\columnwidth]{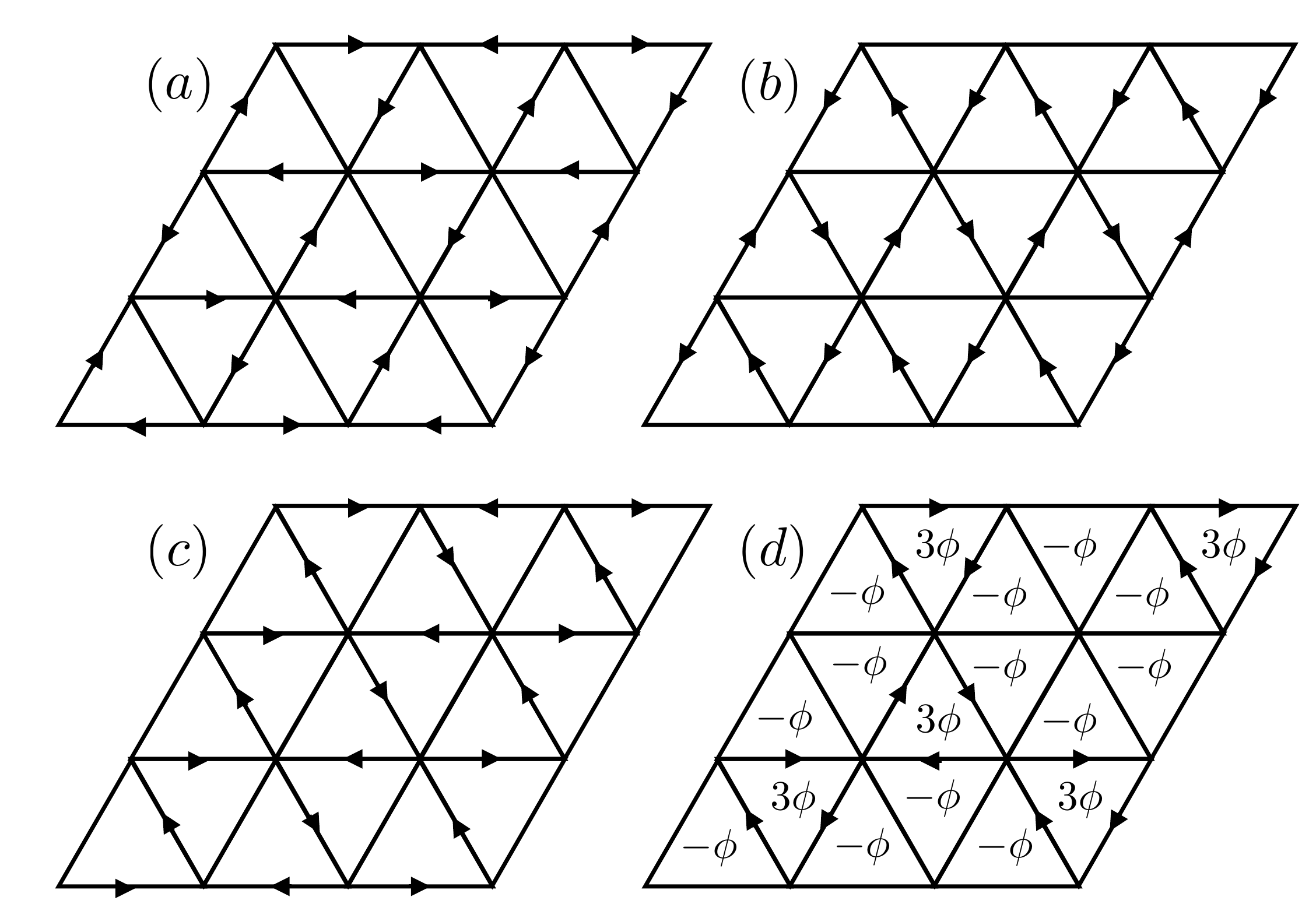}
\caption{\label{fig:trif2}  Real space picture of triangular lattice flux order with $F_2$ symmetry. (a)--(c) Show the flux pattern in terms of imaginary hoppings (i.e., black directed arrows) of the individual components $\vet{M}_\mu$. (d) Shows triple-$M$ order with plaquette fluxes explicitly indicated. We take clockwise arrows to correspond to positive flux. }
\end{figure}

Then, let us consider flux order with $F_2$ symmetry. $F_2$ symmetry implies that form factors associated with each ordering wave vector $\vet{M}_\mu$ break the reflections of the little group, but preserve the two-fold rotation. In case of $\vet{M}_1$ these requirements uniquely determine $\Delta_1(\vet{k})$ and lead to the $d$-wave function $\cos k_3 - \cos k_1 $. It follows from the argument outlined in Appendix~\ref{ssec:compat} that $\Delta_1(\vet{k})$ must be imaginary, which is consistent with flux order. We therefore find for flux order,
\begin{gather} 
\Delta_1(\vet{k})  =  i \Delta_1  (\cos k_3 - \cos k_1 ), \nonumber \\
\Delta_2(\vet{k})  =  i \Delta_2  (\cos k_1 - \cos k_2 ) , \nonumber \\
\Delta_3(\vet{k})  =  i \Delta_3  (\cos k_2 - \cos k_3 ) , \label{eq:f2tri}
\end{gather}
where the $\Delta_\mu$ are real. These $M$-point $d$ waves should be compared to the square lattice $d_{x^2-y^2}$ wave. In case of hexagonal symmetry, there is a $d$ wave for each inequivalent $M$-point component. We found in the previous section that triple-$M$ ordering has $A_2$ symmetry and corresponds to $\Delta_1=\Delta_2=\Delta_3$. The real-space picture of triple-$M$ order in terms of fluxes is shown in Fig.~\eqref{fig:trif2}(d).  

\begin{figure}
\includegraphics[width=\columnwidth]{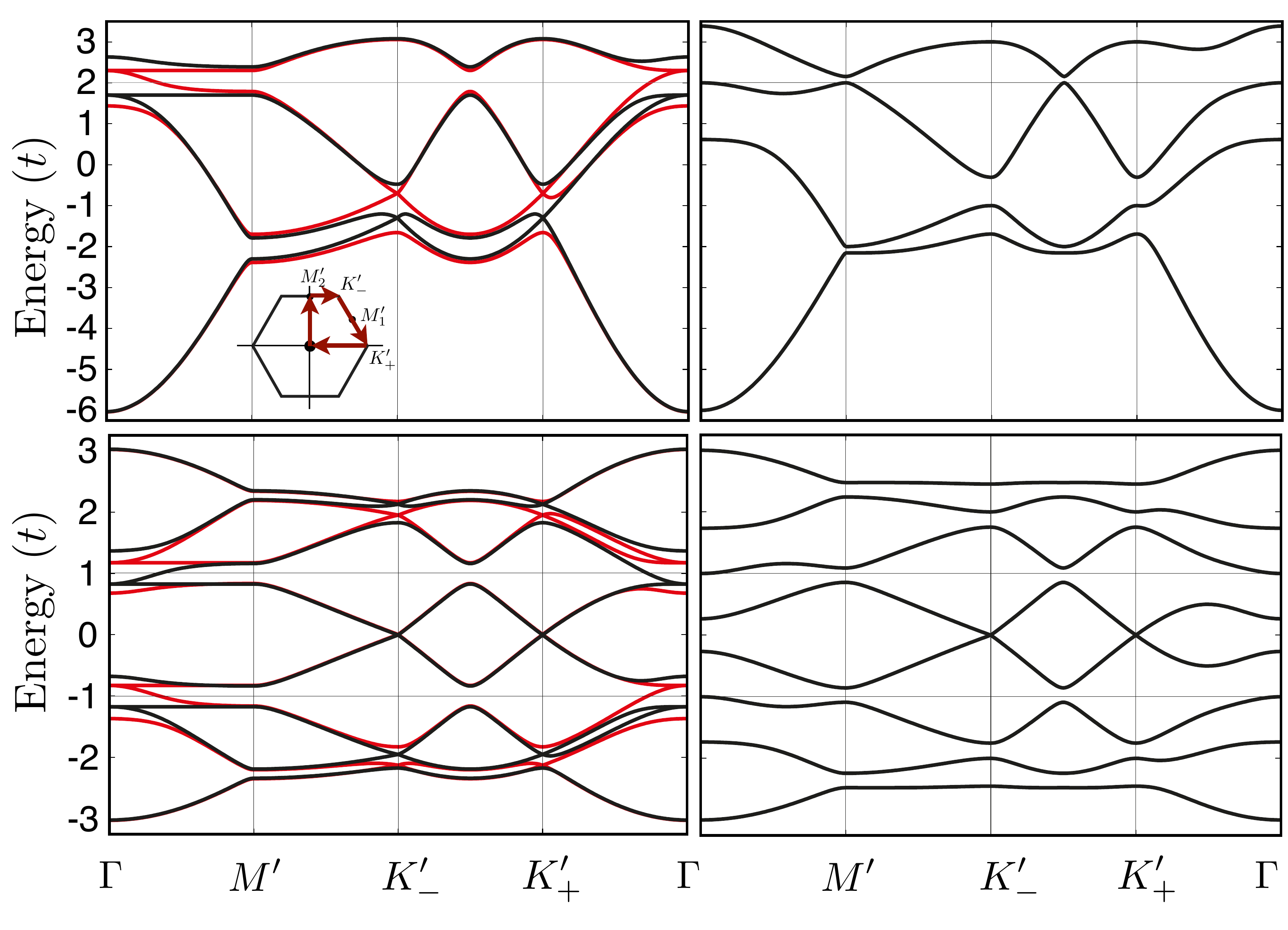}
\caption{\label{fig:spectra} Mean-field spectra of the triple-$M$ charge density ($s$-wave) order (left) and charge-current density ($d$-wave) order (right) of the triangular (top) and and honeycomb (bottom) lattices, given in Eqs.~\eqref{eq:f1tri}, \eqref{eq:f2tri}, \eqref{eq:f1honey}, and \eqref{eq:f2honey}. We show spectra for $\Delta = 0.4$ in case of the triangular lattice $d$-wave state (right top), and $|\Delta|=0.3$ in the other cases. The inset shows the reduced BZ with the path along which bands are plotted. In the left panels the red spectra correspond to $\Delta = -0.3$, whereas black corresponds to $\Delta=0.3$.}
\end{figure}

The mean-field spectra of triangular lattice triple-$M$ $F_1$ (charge density wave) order and $F_2$ flux order states are shown in Fig.~\ref{fig:spectra}. All spectra were calculated for a tight-binding mean-field Hamiltonian $H_0 + H_\Delta$, where $H_0$ contains a nearest-neighbor hopping $t=1$ and $H_\Delta$ contains the mean-fields defined above. In case of charge density wave order we observe that spectrum at $n=3/4$ filling is either gapped (black bands) or semimetallic (red bands), depending on the sign of $\Delta \equiv \Delta_1=\Delta_2=\Delta_3$. The gapless semimetallic point is located at $\Gamma$ of the reduced BZ. The low-energy electronic structure (gapped or gapless) will be the focus of the next section. The spectrum of triple-$M$ $F_2$ flux order is fully gapped irrespective of the sign of $\Delta$. No degeneracies exist at any of the high symmetry points of the folded zone. Given the existence of an energy gap, broken time-reversal invariance, and $A_2$ symmetry we can determine the Chern invariant. To quickly determine whether the Chern number is nonzero, we use eigenvalues of rotation symmetry operators~\cite{fang12} and find that the Chern number is indeed nonzero. We conclude that triple-$M$ $d$-wave order is a Chern insulator state.

\begin{figure}
\includegraphics[width=0.9\columnwidth]{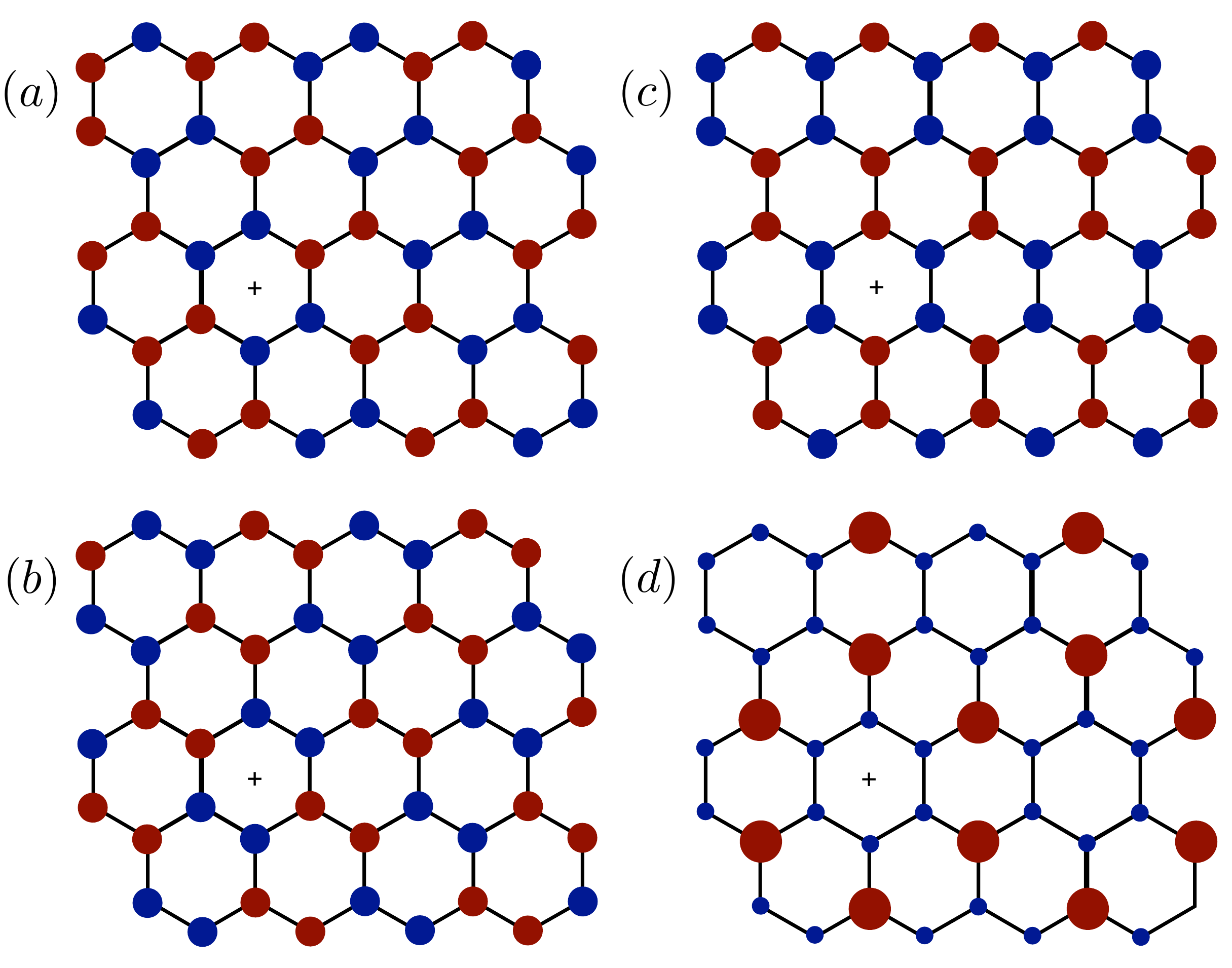}
\caption{\label{fig:honf1site} Real space picture of honeycomb lattice site order with $F_1$ symmetry. (a)--(c) Show the individual components as in Fig.~\ref{fig:trif1} and (d) shows triple-$M$ order.}
\end{figure}

Similar to the triangular lattice, the honeycomb lattice supports site and bond order with $F_1$ symmetry. The honeycomb lattice, however, has two sublattices and in order to express $M$-point site order we make use of two vectors $\vet{w}_A$ and $\vet{w}_B$ which contain the order parameter components for each sublattice. For $F_1$ site order we find that $\vet{w}_A = (-1,-1,1)$ and $\vet{w}_B = (1,-1,-1)$ (for a sketch of how to obtain them see Appendix~\ref{app:mreal}). In terms of these vectors, the condensate functions of $F_1$ site order simply read as
\begin{gather} 
\order{\crea{i \sigma }{\vet{k}+\vet{M}_\mu} \anni{j \sigma' }{\vet{k}}} = \Delta_\mu  w^\mu_i \delta_{ij}\delta_{\sigma\sigma'} ,
\label{eq:f1honey} 
\end{gather} 
where on the right-hand side the index $\mu$ is not summed. Note that the $\Delta_\mu$ are all real. The three individual site order components are shown in Figs.~\ref{fig:honf1site}(a)-\ref{fig:honf1site}(c), which clearly show the resemblance to triangular lattice $F_1$ order. Triple-$M$ order, corresponding to $\Delta_1=\Delta_2=\Delta_3$, is shown in Fig.~\ref{fig:honf1site}(d). 
 
\begin{figure}
\includegraphics[width=\columnwidth]{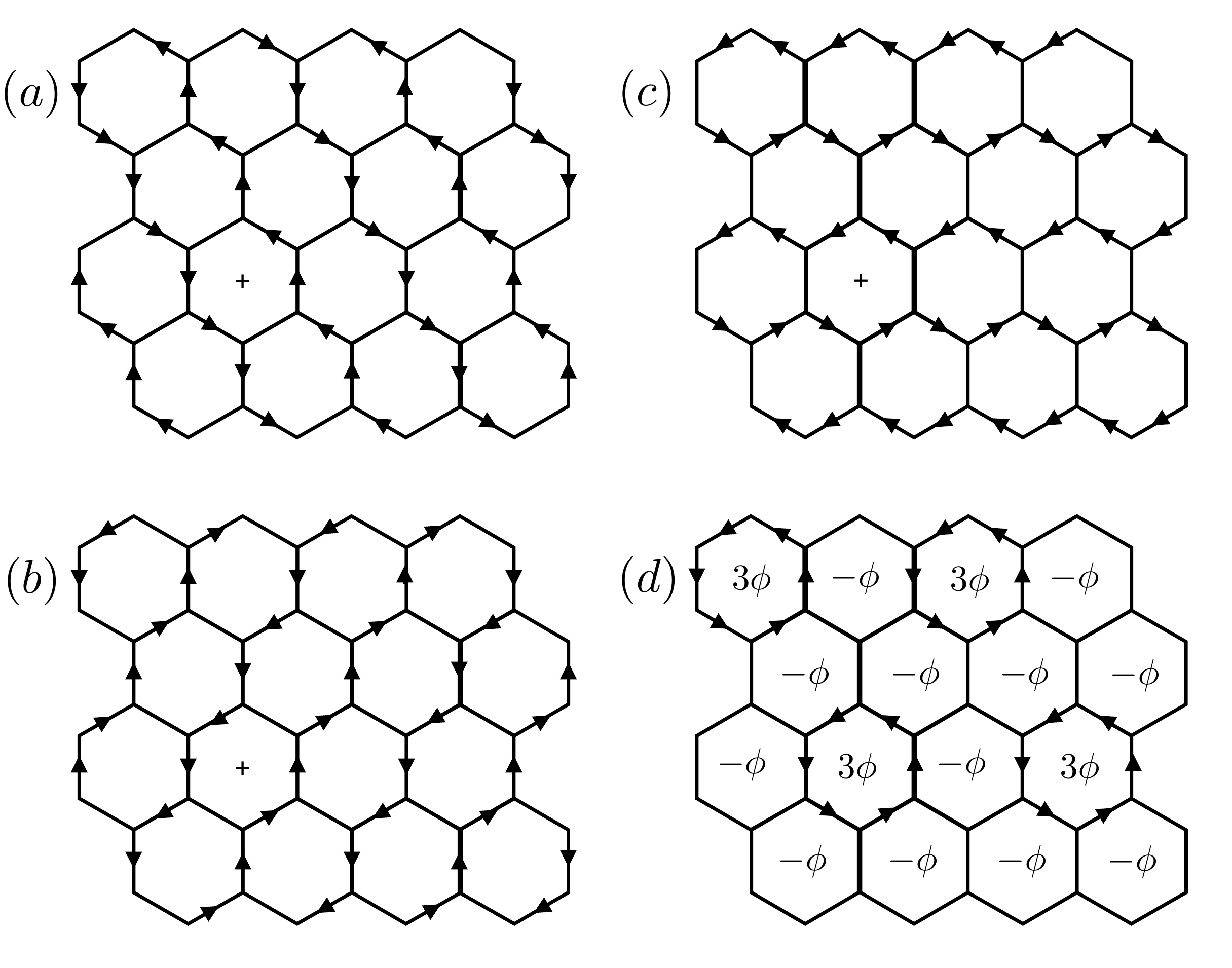}
\caption{\label{fig:honf2} Real space picture of the honeycomb lattice flux order with $F_2$ symmetry. Figures (a)--(c) show the individual components as in Fig.~\ref{fig:trif2} and (d) shows triple-$M$ order. The pattern of fluxes is a direct analog of the triangular lattice as shown in Fig. ~\ref{fig:trif2}(d). }
\end{figure}

We conclude by considering honeycomb lattice flux order with $F_2$ symmetry. Again, due to~\eqref{eq:hermitian} it can be specified by three functions $\Delta_{\mu}(\vet{k}) $. The condensate functions $\Delta_{\mu}(\vet{k}) $ must be chosen so that they are odd under all reflections leaving $\vet{M}_\mu$ invariant. We find that they are given by
\begin{align} 
\Delta_{1}(\vet{k}) =& i\Delta_{1}( e^{-i\delta_1\cdot \vet{k}} - e^{-i\delta_3\cdot \vet{k}} )e^{i\varphi(\vet{k})}, \nonumber \\
\Delta_{2}(\vet{k}) =& i\Delta_{2}( e^{-i\delta_3\cdot \vet{k}} - e^{-i\delta_2\cdot \vet{k}} )e^{i\varphi(\vet{k})}, \nonumber \\
\Delta_{3}(\vet{k}) =& i\Delta_{3}( e^{-i\delta_1\cdot \vet{k}} - e^{-i\delta_2\cdot \vet{k}} )e^{i\varphi(\vet{k})}. \label{eq:f2honey}
\end{align}
Here, $\vec{\delta}_i$ are the vectors connecting nearest neighbors, and we have included a gauge factor $ e^{i\varphi(\vet{k})} \equiv e^{i k_1} $ to enforce our gauge choice (see Appendix~\ref{app:latsym} and Table~\ref{tab:honeyfunctions} of Appendix~\ref{app:gt}). Time-reversal symmetry breaking forces the order parameters $\Delta_\mu$, as defined by Eq.~\eqref{eq:f2honey}, to be real. The three flux order components are shown in Figs.~\ref{fig:honf2}(a)-\ref{fig:honf2}(c), from which it is easily seen that they are honeycomb lattice analogs of Figs.~\ref{fig:trif2}. Specifically, triple-$M$ order, shown in Fig.~\ref{fig:honf2}(d), exhibits the same pattern of fluxes. 

The mean-field spectra of the honeycomb lattice triple-$M$ site order and flux order states are shown in the bottom row of Fig.~\ref{fig:spectra}. We use a honeycomb tight-binding mean-field Hamiltonian $H_0 + H_\Delta$, again with nearest neighbor hopping $t=1$. For both types of order, the spectra closely resemble their triangular lattice analogs. In case of triple-$M$ site order, the spectrum depends on the sign of $\Delta$ ($\equiv \Delta_1=\Delta_2=\Delta_3$), leading to either a gapped or semi-metallic state. Triple-$M$ flux order, on the contrary, leads to a gapped spectrum, and as in case of the triangular lattice we find the mean-field ground state to be a Chern insulator at $n=3/8$. 

In the next section we analyze the spectral properties of mean-field states presented here in more detail. We will do so using a description in terms of low-energy $M$-point electrons, and show that the key features of these states can be understood from simple symmetry considerations. 


\section{Nesting instabilities and properties of mean-field states\label{sec:lowenergy}}

The aim of this section is to develop a deeper understanding of the characteristics of the hexagonal density waves considered in the previous section. To this end, we take a more general perspective and focus on the (low-energy) van Hove electrons: the three flavors of electrons originating from the van Hove singularities located at the $M$-point momenta. Given these three-flavor electrons with relative momenta $\vet{M}_\mu$ we ask which density waves correspond to nesting instabilities. This can be established in a rather straightforward way by deriving the transformation properties of the van Hove electrons under extended point group symmetry and classifying all bilinears according to this symmetry. We will address the gap structures that can arise as a result of density wave formation. The effective filling fraction is $2/3$ for the van Hove electrons, and we find that point node degeneracies protected by symmetry and fully gapped states are possible. The latter come in two varieties: trivial and topological. This will lead us to the general SU(3) theory of $3$-flavor $M$-point electrons applicable to hexagonal lattices. 

Before coming to the hexagonal SU(3) theory, we first review the corresponding SU(2) theory applicable to the square lattice. This is useful in two ways: it will serve to illustrate the general principles at work, and at the same time highlight the difference with the SU(3) theory.

\subsection{SU(2) theory of square lattice van Hove electrons\label{ssec:nestingsq}}

The SU(2) theory of square lattice van Hove electrons is formulated in terms of two flavors of fermions corresponding to the van Hove singularities located at the momenta $\vet{X}$ and $\vet{Y}$. This is schematically shown in Fig.~\ref{fig:bz_square}.  Let $\hat{\Phi}$ be the two-component vector of van Hove electrons, 
\begin{gather} \label{eq:vanhove}
\hat{\Phi} = \begin{pmatrix} \hat{\psi}(\vet{X}) \\ \hat{\psi}(\vet{Y})  \end{pmatrix}.
\end{gather}
The van Hove electrons are connected by the wave vector $\vet{Q}$ (i.e., ``nested''), and a systematic way to address the nesting instabilities is to consider the particle-hole bilinears constructed from them~\cite{markiewicz98}. Specifically, the bilinears $\hat{\Phi}^\dagger  \tau^i  \hat{\Phi}$, where $\tau^i$ are Pauli matrices, define the algebra of the van Hove singularities, and $\tau^3$ measures the population imbalance of the two species of van Hove electrons. All terms that do not commute with $\tau^3$ lead to nesting instabilities. 
These terms are simply given by $\tau^1$ and $\tau^2$. To identify which states they physically correspond to, we derive how they transform under the symmetry operations of the lattice. We find that the action of symmetry group generators on $\hat{\Phi}$ is given by (see Appendix~\ref{ssec:symtsb})
\begin{align} 
T(\vet{a}_1) \; : \; \hat{\Phi} \; & \rightarrow  \;  -\tau^3  \hat{\Phi} \nonumber \\
C_4 \;  : \; \hat{\Phi} \; & \rightarrow \;   -i \tau^2 \hat{\Phi}\nonumber \\
\sigma_v  \;  : \; \hat{\Phi} \; & \rightarrow \;  \tau^3  \hat{\Phi} \label{eq:symvanhove}
\end{align}
These transformation properties, in combination with time reversal $\Theta$ which acts as complex conjugation, show that $\tau^1$ and $\tau^2$ have $B'_2$ and $A'_2$ symmetry, respectively. Going back to the decompositions of Eqs.~\eqref{eq:sqsite} and~\eqref{eq:sqflux} we simply find the familiar result that the staggered charge density wave ($B'_2$ symmetry) and $d_{x^2-y^2}$-density wave ($A'_2$ symmetry) are the nesting instabilities. We note in passing that it is straightforward to include the electron spin, in which case $\hat{\Phi}^\dagger  \tau^1\sigma^i  \hat{\Phi}$ corresponds to the antiferromagnetic spin density wave instability. 

Clearly, the two bilinears $\hat{\Phi}^\dagger  \tau^1  \hat{\Phi}$ and $\hat{\Phi}^\dagger  \tau^2  \hat{\Phi}$ lift the degeneracy of the van Hove electrons and may be viewed as mass terms in the low-energy subspace $\hat{\Phi} $. This does not, however, necessarily imply a fully gapped mean-field spectrum. For instance, the $d_{x^2-y^2}$ form factor has nodal lines which intersect the Fermi surface, and as a result the $d_{x^2-y^2}$-density wave has point nodes at the boundary of the reduced BZ. In contrast, the charge density wave is an $s$-wave condensate and does not have nodes. The nodal degeneracies of the $d_{x^2-y^2}$-density wave are protected by lattice symmetries, which can be established in a direct and systematic way using the extended point group $C'''_{4v}$ and its representations. 

The nodes of the $d_{x^2-y^2}$-density wave are located at the two inequivalent momenta $\vet{q}_0= \pi(1,1)/2a$ and $\vet{q}'_0= \pi(-1,1)/2a$. The node at $\vet{q}_0$ is shown in Fig.~\ref{fig:bz_square}. To study the symmetry protection of the nodal degeneracies, we focus on $\vet{q}_0$ and abbreviate the fermion operator at $\vet{k}=\vet{q}_0$ as 
\begin{gather} 
\hat{\chi}(\vet{q}_0) = \begin{pmatrix} \hat{\chi}_{0}(\vet{q}_0) \\ \hat{\chi}_{1}(\vet{q}_0)  \end{pmatrix} = \begin{pmatrix} \hat{\psi}(\vet{q}_0) \\ \hat{\psi}(\vet{q}_0+\vet{Q})  \end{pmatrix}.
\end{gather}
The symmetry of the $d_{x^2-y^2}$-density wave is $A'_2$, and from the character table of $C'''_{4v}$ we read off the invariant elements. A subset of these symmetries leave $\vet{q}_0$ invariant and can be used to derive constraints on the coupling between the two degenerate states at $\vet{q}_0$. Two of such symmetries are the inversion $C_2$, and the combination of the reflection $\sigma_{1d} = C_4\sigma_v$ and $T(\vet{a}_1)$. Applying Eq.~\eqref{eq:tsb} of Appendix~\ref{ssec:symtsb}, we find that these symmetries act on $\hat{\Phi}$ as
\begin{align}
C_2   \;  : \; \hat{\chi}(\vet{q}_0) \; &  \rightarrow \; \hat{\chi}(-\vet{q}_0)= \hat{\chi}(\vet{q}_0-\vet{Q}) =  \tau^1 \hat{\chi}(\vet{q}_0) \nonumber \\
T(\vet{a}_1)\sigma_{1d}  \;  : \; \hat{\chi}(\vet{q}_0)\; &  \rightarrow \; \tau^3\hat{\chi}(\vet{q}_0),
\end{align}
where $\tau^i$ is a set of Pauli matrices. At $\vet{q}_0$, the mean-field Hamiltonian must commute with both $\tau^1$ and $\tau^3$ and the only matrix which has this property is the unit matrix. It follows that these symmetries protect the degeneracy at $\vet{q}_0$. 

It is possible to show that a single symmetry protects the degeneracy at $\vet{q}_0$. Time-reversal symmetry $\Theta$ and the translation $T(\vet{a}_1)$ are broken but their product is preserved and leaves $\vet{q}_0$ invariant in the reduced BZ. The action of $\Theta T(\vet{a}_1)$ on $\Phi$ is
 \begin{gather}
\Theta T(\vet{a}_1) \;  : \; \hat{\Phi}  \; \rightarrow \; \mathcal{K}\tau^3\tau^1 \hat{\Phi} ,
\end{gather}
with $\mathcal{K}$ complex conjugation. This leads to the condition $\tau^3\tau^1 \mathcal{H}^*(\vet{q}_0)\tau^1\tau^3 = \mathcal{H}(\vet{q}_0)$, which forces $\mathcal{H}(\vet{q}_0)$ to be proportional to the identity and proves that the degeneracy is protected by the symmetry. 

The $d$-wave orbitals $d_{x^2-y^2}$ and $d_{xy}$ are not degenerate in the presence of square symmetry (i.e., point group $C_{4v}$). As a consequence, the $d_{x^2-y^2}$- and $d_{xy}$-density waves are different in nature: the former constitutes a nesting instability, whereas the latter does not. In particular, this implies that projecting the $d_{xy}$-density wave into the subspace of van Hove electrons~\eqref{eq:vanhove} simply gives the identity: no mass term is generated since the nodal lines of the $d_{xy}$ form factor cross the van Hove electrons, which remain degenerate. It is a simple matter to show that this degeneracy is protected by symmetry. The fourfold rotation $C_4$ acts as $\sim \tau^2$ and time-reversal symmetry acts as complex conjugation $\mathcal{K}$, implying that the combination of both symmetries protects the degeneracy of $\hat{\Phi}$. 

An alternative and illuminating way to address the spectral properties of the $d_{xy}$-density wave is to expand the mean-field spectrum in small momenta $\vet{q}$ in the subspace $\hat{\Phi}(\vet{q})$. Such an expansion yields the low-energy Hamiltonian 
\begin{gather} \label{eq:hamQBC}
\mathcal{H}(\vet{q}) =  2t(q^2_x - q^2_y)\tau^3 -2  \Delta_{d_{xy}} 2 q_xq_y\tau^1,
\end{gather}
where $ \Delta_{d_{xy}}$ denotes the amplitude of the $d_{xy}$-density wave. We recognize that this low-energy Hamiltonian takes the form of a quadratic band crossing (QBC) Hamiltonian. QBCs are known to be protected by symmetry~\cite{sun09}. A spectral gap at $\vet{q}=0$ would correspond to a term proportional to $\tau^2$, the appearance of which is forbidden by time-reversal symmetry. Momentum independent terms proportional to $\tau^3$ and $\tau^1$ are odd under the fourfold rotation, and it follows that the quadratic band crossing degeneracy is protected by $C_4$ symmetry. Note that the quadratic band crossing Hamiltonian only arises in the presence of the $d_{xy}$-density wave. Indeed, the transformation properties of $\hat{\Phi}$ given in~\eqref{eq:symvanhove} imply that the $ \Delta_{d_{xy}} $ term is odd under translations. 

A QBC may be gapped out or split into two Dirac cones~\cite{sun09}. From our symmetry analysis we directly find the terms which have this effect. The time-reversal odd term $\tau^2$ which gaps the QBC simply corresponds to staggered flux order of $A'_2$ symmetry, i.e., the $d_{x^2-y^2}$-density wave. This is consistent with the statement that the $d+id$ state is gapped and equivalent to a quantum Hall state. We found that $\tau^1$ has $B'_2$ symmetry and corresponds to the charge density wave. The term $\tau^3$ has $B_1$ symmetry and according to Eq.~\eqref{eq:sqbond} is induced by translational invariant bond order. In the language of Ref.~\onlinecite{sun09} the $C_4$ breaking terms $\tau^1$ and $\tau^3$ are nematic site and bond order, respectively. 

We can ask a similar question for the nodes of the $d_{x^2-y^2}$-density wave: Which symmetry breaking perturbations shift or gap out the degeneracies? The nodes of the $d_{x^2-y^2}$ state have linear Dirac dispersion and a symmetry analysis of the low-energy Dirac theory, specifically of Dirac fermion bilinears, provides a systematic answer to this question. For instance, based on the principle of reciprocity we conclude that the $A'_1$ state, the $d_{xy}$-density wave, produces a Haldane gap with a Chern-insulating ground state. The bond ordered states of $E_3$ symmetry are the square lattice counterparts of the Kekul\'e bond orderings on the honeycomb lattice. They couple to Dirac fermions mass bilinears. This may be understood from the decomposition of the representation $E_3$ into representations of the point group $C_{4v}$ given by $E_3 \rightarrow A_1 + B_1$. In contrast, the doublets of $E'_1$ and $E_5$ symmetry lead to an effective pseudo-gauge-field coupling in the Dirac theory, since both representations project onto the $E_1$ representation of $C_{4v}$. 

Revisiting the square lattice density waves highlights the utility of the systematic framework provided by the extended point group symmetries and representations. Nesting instabilities can be directly associated with density wave states through symmetry, and low-energy properties, in particular spectral degeneracies, can be determined from the action of extended point group symmetries. Importantly, the latter automatically take the effect of fractional translations, i.e., translations which are not symmetries by themselves, into account. In fact, since the extended point groups take the effect of such fractional translations into account, they can be used to describe materials with non-symmorphic symmetries, in cases where the non-symmorphic symmetries result from translational symmetry breaking. The presence of non-symmorphic symmetries can give rise to protected Dirac semimetals in three and two dimensions~\cite{young12,young15}. Consequently, extended point groups will find application in the study of symmetry-protected semimetals.

\subsection{SU(3) theory of $M$-point electrons \label{ssec:nestinghexa}}

\begin{figure}
\includegraphics[width=\columnwidth]{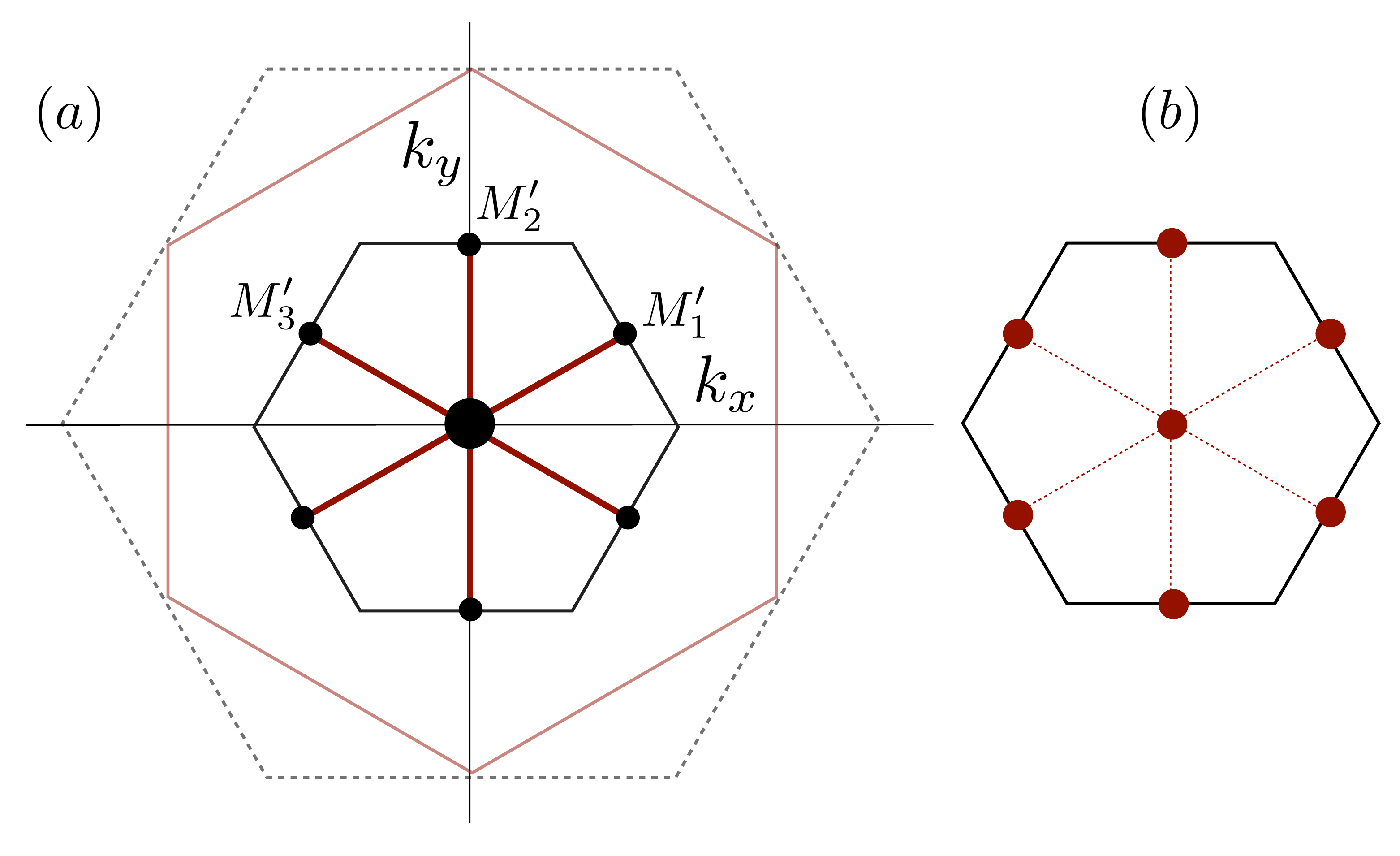}
\caption{\label{fig:foldedzone} (a) Folded Brillouin zone (BZ) due to quadrupling of unit cell (smaller solid black hexagon). Red lines in the folded BZ represent the folded Fermi surface. The Fermi surface lines are doubly degenerate everywhere, except for their crossing at $\Gamma$. The two dashed hexagons are the unfolded BZ (black) and the Fermi surface (red). (b) Folded BZ high-symmetry points coinciding with the Fermi surface, i.e., $\Gamma$ and $M'$.}
\end{figure}

The analysis of hexagonal lattice nesting instabilities is a generalization of the analysis for the square lattice. The difference with respect to the latter is that instead of two there are three van Hove singularities, located at $\vet{M}_\mu$, which are mutually connected by three inequivalent wave vectors, also given by $\vet{M}_\mu$. This gives rise to three flavors of van Hove electrons labeled by $\mu=1,2,3$. At the corresponding filling $n$ (e.g., $n=3/4$ for the triangular lattice; $n=3/8$ for the honeycomb lattice) the Fermi surface of the simple nearest-neighbor tight-binding band structure is a hexagon with vertices at $\vet{M}_\mu$, shown in Fig.~\ref{fig:bz_hexa}. In Fig.~\ref{fig:foldedzone} we show the folding of the perfectly nested Fermi surface into the reduced Brillouin zone. The existence of the van Hove singularities and their algebra do not depend on the ideal situation of perfect nesting. 

The algebra of the three van Hove singularities is expressed in terms of three flavors of fermions $\hat{\psi}_\mu$, collected in the van Hove operator $\hat{\Phi} $,
\begin{gather} \label{eq:vanhovehexa}
\hat{\Phi} =\begin{pmatrix} \hat{\psi}_1 \\ \hat{\psi}_2  \\ \hat{\psi}_3 \end{pmatrix}=  \begin{pmatrix} \anni{ }{\vet{M}_1} \\ \anni{}{\vet{M}_2}   \\ \anni{ }{\vet{M}_3}   \end{pmatrix}.
\end{gather}
The full algebra is given by the bilinears $\hat{\Phi}^\dagger \Lambda^i \hat{\Phi}$, where (in contrast to the $\tau^i$) the matrices $\Lambda^i $ are the generators of $SU(3)$, i.e., the Gell-Mann matrices and the identity. It is convenient to group the eight Gell-Mann matrices in three distinct sets: two sets of three, $\vet{\Lambda}_a\equiv  (\Lambda^1_a,\Lambda^2_a,\Lambda^3_a)$ which is explicitly given by
\begin{gather}
\Lambda^1_{a} =  \begin{pmatrix} 0 & 1 & 0 \\ 1 & 0& 0 \\ 0&0&0 \end{pmatrix}, \; \Lambda^2_{a} = \begin{pmatrix} 0 & 0 & 0 \\ 0 & 0& 1 \\ 0 & 1&0 \end{pmatrix}, \; \Lambda^3_{a} = \begin{pmatrix} 0 & 0 & 1 \\ 0 & 0& 0 \\ 1 & 0&0 \end{pmatrix}, \label{eq:gellmanna}
\end{gather}
and $\vet{\Lambda}_b\equiv (\Lambda^1_b,\Lambda^2_b,\Lambda^3_b)$ given by
\begin{gather}
\Lambda^1_{b} =  \begin{pmatrix} 0 & -i & 0 \\ i & 0& 0 \\ 0&0&0 \end{pmatrix}, \; \Lambda^2_{b} = \begin{pmatrix} 0 & 0 & 0 \\ 0 & 0& -i \\ 0 & i&0 \end{pmatrix} , \; \Lambda^3_{b} = \begin{pmatrix} 0 & 0 & i \\ 0 & 0& 0 \\ - i & 0&0 \end{pmatrix} . \label{eq:gellmannb}
\end{gather}
A third set of two, $\vet{\Lambda}_c\equiv (\Lambda^1_c,\Lambda^2_c)$, is given by
\begin{gather}
\Lambda^1_{c} =\begin{pmatrix} 1 & 0 & 0 \\ 0 & -1 & 0 \\ 0&0&0 \end{pmatrix}, \; \Lambda^2_{c} =   \frac{1}{\sqrt{3}} \begin{pmatrix} 1 & 0 & 0 \\ 0 & 1 & 0 \\ 0 & 0& -2 \end{pmatrix}   . \label{eq:gellmannc}
\end{gather}
An important property of the Gell-Mann matrices is that they give rise to three distinct SU(2) subalgebras, which correspond to the three different ways of coupling the van Hove electrons. For instance, one of such subalgebras is given by the matrices $(\Lambda^1_a,\Lambda^1_b,\Lambda^1_c)$, all of which act in the subspace of $(  \hat{\psi}_1 ,\hat{\psi}_2  )$ and therefore couple $\vet{M}_1$ and $\vet{M}_2$. As a consequence, the algebra generated by this triple of Gell-Mann matrices governs the nesting instabilities at wave vector $\vet{M}_3$, the vector connecting $\vet{M}_1$ and $\vet{M}_2$. In analogy with the square lattice, the matrix $\Lambda^1_c$ is interpreted as the population imbalance between van Hove electrons $ \hat{\psi}_1 $ and $\hat{\psi}_2 $, and the two non-commuting matrices $\Lambda^1_a$ and $\Lambda^1_b$ constitute the nesting instabilities. The nesting instabilities at wave vectors $\vet{M}_1$ and $\vet{M}_2$ are obtained by either explicitly forming the appropriate subalgebras acting in the subspace of $(  \hat{\psi}_2 ,\hat{\psi}_3  )$ and $(  \hat{\psi}_3 ,\hat{\psi}_1  )$, respectively, or more directly by invoking rotational symmetry. 

Now that we have obtained the nesting instabilities we proceed to discuss three aspects of the low-energy van Hove electrons: \emph{(i)} degeneracies and splitting of energy levels, \emph{(ii)} the symmetry of the nesting instabilities and their identification with density wave order, and \emph{(iii)} the properties of triple-$M$ ordering.
To this end we first ask how the generators of the extended symmetry group $C'''_{6v}$ act on the van Hove electrons $\hat{\Phi}$ given in~\eqref{eq:vanhovehexa}. We find that the action is 
\begin{align}
T(\vet{a}_1) \; : \;  \hat{\Phi} &\rightarrow \; G_1 \hat{\Phi}, \nonumber \\
C_6 \; : \;  \hat{\Phi} &\rightarrow \; X \hat{\Phi} \nonumber \\
\sigma_v \; : \;  \hat{\Phi} &\rightarrow \; Y \hat{\Phi} \label{eq:mrepvanhove}
\end{align}
The matrices $G_1$, $X $, and $Y$ define a three-dimensional matrix representation. Their explicit expressions and a more detailed discussion of this $M$-point representation is given in Appendix~\ref{app:malgebra}. In the presence of full $C'''_{6v}$ symmetry, i.e., without symmetry breaking, the representation is irreducible and given by $F_1$. This enforces a trivial three-fold degeneracy of energies at the $M$-points, which is lifted when symmetry is lowered. For instance, when only translational symmetry is lost, the symmetry group is $C_{6v}$ and the representation $F_1$ (now reducible) is decomposed as $A_1+ E_2$. If the symmetry is lowered to $C_{3v}$, the decomposition is $A_1 + E $~\cite{note1}. We see that a two-fold degeneracy persists down to rhombohedral $C_{3v}$ symmetry. In contrast, when the symmetry is lowered to hexagonal $C_6$ or orthorhombic $C_{2v}$, the lack of two-dimensional representations implies the absence of protected degeneracies.

This analysis can be directly applied to the symmetry breaking as a result of electronic ordering corresponding to the nesting instabilities. First, we establish the symmetry of the nesting instabilities. Using the action of the symmetry group expressed in Eq.~\eqref{eq:mrepvanhove}, we find that the set of matrices $\vet{\Lambda}_a$ transforms as $F_1$ and the set $\vet{\Lambda}_b$ transforms as $F_2$. Therefore, the former has the same symmetry as the charge density $s$ waves, whereas the latter has the same symmetry as the flux density $d$ waves, both of which we studied in previous sections. We thus find that $\vet{\Lambda}_a$ describes $s$-wave particle-hole instabilities and $\vet{\Lambda}_b$ describes $d$-wave particle-hole instabilities. This is consistent with $\vet{\Lambda}_b$ being time-reversal odd. 

\begin{figure}
\includegraphics[width=\columnwidth]{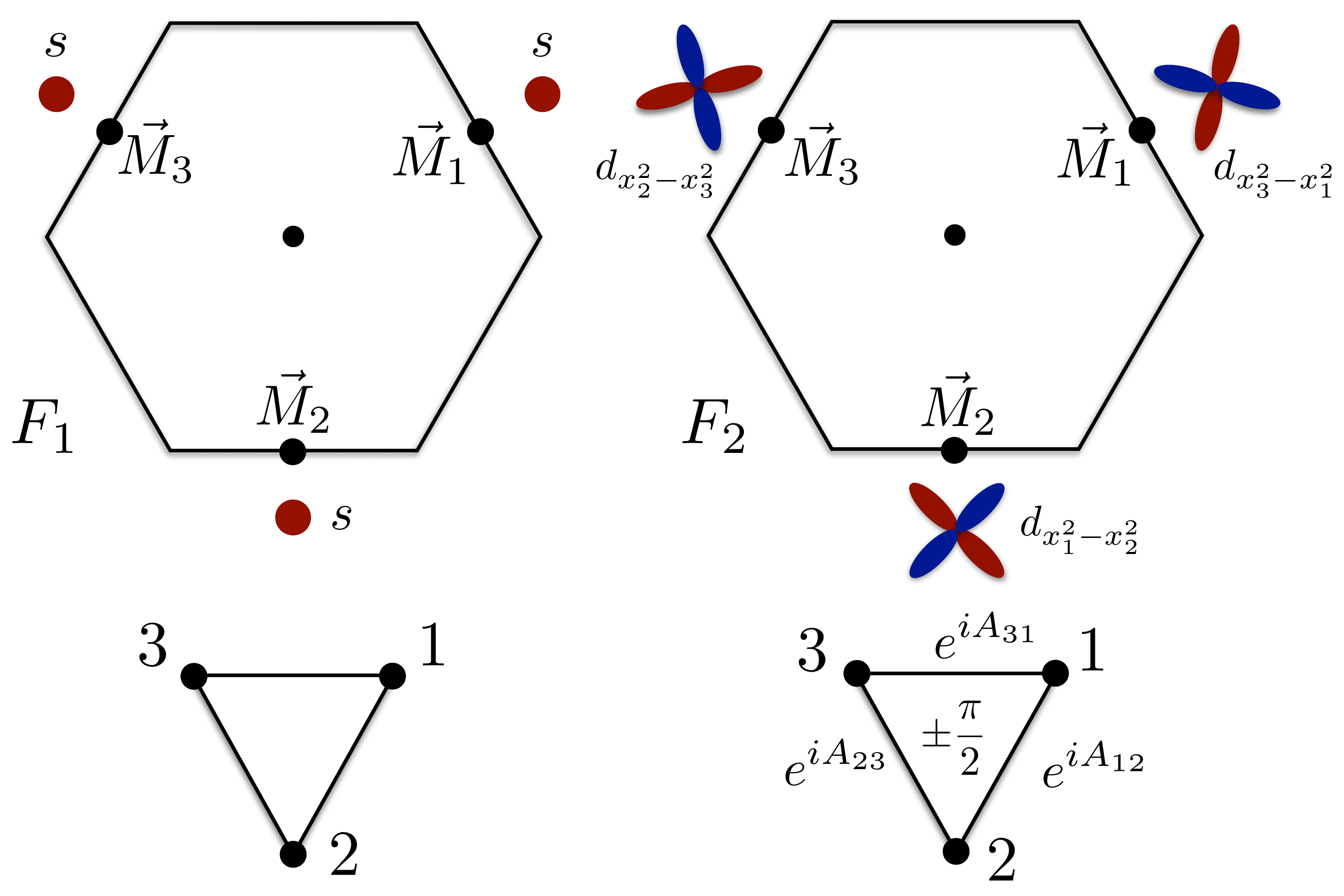}
\caption{\label{fig:mpointorder} (Top) Schematic representation of hexagonal triple-$M$ ordering: (top left) an orbital with $s$-wave symmetry is associated with each $M$ point in case of charge density wave order, whereas (top right) an orbital with $d$-wave symmetry is associated with each $M$ point in case of flux (or charge-current) order. The $d$-wave orbitals are rotated by $\pm 2\pi/3$ with respect to each other. (Bottom) In both cases, triple-$M$ ordering is formally equivalent to electrons hopping on a three-site triangular loop, where in case of $d$-wave ordering (bottom right) a flux of $\pm \pi/2$ is pierced through the loop.}
\end{figure}

Now, based on the identification of the nesting instabilities, let us consider triple-$M$ ordering, i.e., simultaneous ordering at the three wave vectors $\vet{M}_\mu$. Since the charge density $s$ waves have $F_1$ symmetry, triple-$M$ $s$-wave ordering has $A_1$ symmetry. In contrast, flux density $d$-wave order has $F_2$ symmetry, implying $A_2$ symmetry for triple-$M$ $d$-wave order. As a result, the symmetry groups of the ordered states are $C_{6v}$ and $C_6$, respectively. 
We can obtain the energy levels for triple-$M$ order in the subspace of van Hove electrons $\hat{\psi}_\mu$ by considering symmetric sums of the $\Lambda^i$ matrices which describe the order. In particular, triple-$M$ charge density wave order is given by the sum $\Lambda_a = \sum_i \Lambda^i_a$, and similarly for $d$-wave order, i.e., $\Lambda_b = \sum_i \Lambda^i_b$. 
Diagonalization of these two terms shows that the corresponding eigenvalue matrices are given by $\Lambda^2_c \sim \text{diag}(1,1,-2)$ and $\Lambda^2_c \sim \text{diag}(1,-1,0)$. This is consistent with respective $C_{6v}$ and $C_{6}$ symmetry, the former having a protected degeneracy. It is interesting to observe that the triple-$M$ matrices $\Lambda_a $ and $\Lambda_b$, which connect electrons in momentum space, are equivalent to hopping matrices describing electrons hopping on single three-site triangular loop. In case of time-reversal breaking $d$-wave order $\Lambda_b$ the loop is pierced with a U(1) flux of $\pm \pi/2 = A_{12} + A_{23}+ A_{31}$, as is shown in Fig.~\ref{fig:mpointorder}, where the sign depends on the overall sign of the order parameter. 

We further observe that the two eigenvalues matrices coincide with the third set of set of Gell-Mann matrices, $\Lambda^1_c$ and $\Lambda^2_c$. The matrices $\vet{\Lambda}_c = (\Lambda^1_c,\Lambda^2_c)$ may therefore be viewed as encoding the possible energy level splittings due to triple-$M$ order. It is worth mentioning that $\Lambda^1_c$ leads to a symmetric splitting of energies, whereas $\Lambda^2_c$ does not. In the latter case, it remains undetermined whether the degenerate level has higher energy than the non-degenerate level. We come back to this ambiguity towards the end of this section. 

It is important to stress the matrices $\vet{\Lambda}_c $ represent the splitting of energy levels due to triple-$M$ order only in the eigenbasis. In contrast, the bilinears $\hat{\Phi}^\dagger \Lambda^1_c \hat{\Phi}$ and $\hat{\Phi}^\dagger \Lambda^2_c \hat{\Phi}$, which do not carry momentum, actually describe the two $Q=0$ $d$-wave components $(d_{x^2-y^2}, d_{xy})$, e.g., Eqn.~\eqref{eq:e2tri}. Therefore, in the basis of van Hove electrons, the matrices $\vet{\Lambda}_c $ encode the nematic order associated with $Q=0$ $d$-wave order.

Figure~\ref{fig:foldedzone} shows that the Fermi surface also intersects the $M'$ points of the folded Brillouin zone. These degeneracies at high-symmetry points should also be considered. The analysis is straightforward due to the reduced symmetry of the $M'$ points as compared to $\Gamma$. Unless translational symmetry is preserved, we expect the degeneracies to be lifted due to the lack of two-dimensional representations for $C_{2v}$ symmetry. Translations act nontrivially at the zone boundary. We find that the two-fold degeneracy at $M$ is generally lifted. We leave the details to Appendix~\ref{app:mpoint}.

To gain further insight in the structure of hexagonal $M$-point order we expand the kinetic Hamiltonian $H_0$ in small momenta $\vet{q}$ around $\Gamma$ in the basis $\hat{\Phi}(\vet{q}) =[ \hat{\psi}_1(\vet{q}) ,\hat{\psi}_2(\vet{q}) , \hat{\psi}_3(\vet{q}) ]^T $. For definiteness we particularize to the triangular lattice, but equivalent results can be obtained for all hexagonal lattices. To second order in $\vet{q}$ we find, in terms of $q_i = \vet{q}\cdot \vet{a}_i$,
\begin{gather} \label{eq:LEhamM}
\mathcal{H}(\vet{q}) =   \begin{pmatrix} - q_1q_3 & &    \\ & -q_2q_1 &    \\  & & -q_3q_2   \end{pmatrix}.
\end{gather}
We rewrite the Hamiltonian in a basis given by the states $\hat{\Phi}_{A} $ and $\hat{\Phi}_{E} $, which are expressed in terms of the $\hat{\psi}_\mu$ as
\begin{align} \label{eq:Mbasisstates}
\hat{\Phi}_{A}  & =  \frac{1}{\sqrt{3}} ( \hat{\psi}_1 + \hat{\psi}_2 + \hat{\psi}_3 ),  \\
\hat{\Phi}_{E}  & = \left\{  \begin{matrix} \frac{1}{\sqrt{6}} ( \hat{\psi}_{1} + \hat{\psi}_{2} -2 \hat{\psi}_{3}  ), \\ \frac{1}{\sqrt{2}} ( - \hat{\psi}_{1} + \hat{\psi}_{2} ), \end{matrix} \right. 
\end{align}
Applying this basis transformation we find that the Hamiltonian projected into the subspace defined by $\hat{\Phi}_{E} $ takes the form
\begin{gather}  \label{eq:LEhamME}
\mathcal{H}_{E}(\vet{q}) = \pm \frac{1}{4}q^2 \pm \frac{1}{4}\left[ (q_x^2 - q^2_y )\tau^3 + 2q_xq_y\tau^1\right],
\end{gather}
where $\tau^3=\pm 1$ labels the two states. The appearance of the $d$-wave functions $(q_x^2 - q^2_y ,2q_xq_y )$ gives the Hamiltonian $\mathcal{H}_{E}(\vet{q})$ the structure of a QBC point. QBC points are topological semimetallic points that enjoy special potection~\cite{sun09}. This is consistent with our finding that a two-fold degeneracy is protected by point group symmetry: the topological quadratic band touching is the protected two-fold degeneracy. Furthermore, from this identification we infer that the two-fold degeneracy cannot be lifted without breaking time-reversal symmetry. Indeed, a gap opening would be induced by a term proportional to $\tau^2$, which is time-reversal odd. The QBC Hamiltonian of Eq.~\eqref{eq:LEhamME} can be directly compared to the Hamiltonian of Eq.~\eqref{eq:hamQBC} of the square lattice van Hove electrons. There is an important difference between the two. In the hexagonal case, the QBC arises in the fully symmetric low-energy expansion. The reason is the degeneracy of the $d$-wave functions. 

To make a connection with the spectral effect of triple-$M$ order discussed earlier, we note that $s$-wave order splits off the $\hat{\Phi}_{A} $ state from the quadratic band touching doublet $\hat{\Phi}_{E} $ and preserves the degeneracy of the latter. The matrix $\Lambda^2_c$ describes this energy level splitting in this basis. Time-reversal odd $d$-wave order couples the doublet $\hat{\Phi}_{E} $ by introducing the term $\tau^2$, and has no effect on $\hat{\Phi}_{A} $. The coupling leads to a splitting, which is described by $\Lambda^1_c$. The nontrivial nature of the resulting energy gap follows directly from the splitting of a quadratic band touching with $2\pi$ Berry flux~\cite{sun09}.

That such a gap opening leads to a Chern insulating phase may be understood from symmetry. The Chern number can be obtained $\text{mod}\, 3$ by multiplying the $C_3$ rotation eigenvalues of all occupied bands at $C_3$ invariant BZ points~\cite{fang12}. The $C_3$ eigenvalues of $\hat{\Phi}_{E} $ are $e^{\pm i 2\pi/3 }$, while the product of $C_3$ eigenvalues of occupied states at $\vet{K}'_+$ and $\vet{K}'_-$ must be trivial, i.e., equal $1$. This proves that if the $\hat{\Phi}_{E} $ doublet is split and one of the states is occupied, the Chern number will be nonzero.

\begin{table}[t]
\centering
\begin{ruledtabular}
\begin{tabular}{ccccl}
Rep. & Type  &  triple-$M$ &  Ground state  & $\Theta$ \\ 
\hline
$F_1$  & $s$-wave & $A_1$ & insulator/QBC semi-metal  & $+$\\
$F_2$  & $d$-wave & $A_2$  & Chern insulator  &  $-$\\
\end{tabular}
\end{ruledtabular}
 \caption{Table summarizing the key properties of the multi-component hexagonal lattice $s$- and $d$-density wave orders, which correspond to nesting instabilities. The triple-$M$ mean-field ground state has $A_1$ ($A_2$) symmetry, is time-reversal even (odd), and is insulating or a QBC semi-metal (Chern insulator). }
\label{tab:morders}
\end{table}

In the context of $M$-point ordering, the splitting of a quadratic band touching was found to explain the quantum anomalous Hall effect in the ground state of a chiral triple-$M$ spin density wave~\cite{chern12}. Our analysis shows that this scenario is a generic feature of hexagonal lattice $M$-point order. Indeed, the present analysis can be generalized to explicitly take the electron spin into account, leading to a unified framework of $M$-point order~\cite{venderbos}.

We now comment on the energies of the states $\hat{\Phi}_{A} $ and $\hat{\Phi}_{E} $ in case of triple-$M$ $s$-wave order. The energies (more appropriately, the energy shifts) can be written in terms of a triple-$M$ (real) order parameter $\Delta_s$ as $\Delta_s \Lambda^2_c$, and we see that this is not symmetric with respect to $\Delta_s \rightarrow -\Delta_s$. Whether the non-degenerate or degenerate level is higher in energy (i.e., the relative ordering of $\hat{\Phi}_{A} $ and $\hat{\Phi}_{E} $) is determined by the sign of $\Delta_s$. This can be related to a Ginzburg-Landau expansion of the free energy in terms of the order parameter $\Delta_s$, which allows for a cubic term $\sim c \Delta^3_s$~\cite{maharaj13,venderbos}. As a result, the sign of $\alpha$ determines the sign of $\Delta_s$. In contrast, no such term exist for triple-$M$ $d$-wave order $\Delta_d$, in agreement with the fact that the sign of $\Delta_d$ does not affect the spectrum. We expect and find that the sign of $\Delta_s$ is such that the energy of the doublet $\hat{\Phi}_{E} $ is lowered, leading to the opening of a full energy gap.

The main results obtained for hexagonal lattice $M$-point order are summarized in Table~\ref{tab:morders}. We find two nesting instabilities, characterized by $F_1$ ($s$-wave) and $F_2$ ($d$-wave) symmetry, with preserved and broken time-reversal symmetry, respectively. Simultaneous ordering of the three components (i.e., triple-$M$ order), schematically shown in Fig.~\ref{fig:mpointorder}, leads to an insulating mean-field ground state in case of $s$-waves, and to a QAH insulating ground state in case of $d$ waves. 


\section{Discussion and conclusion\label{sec:summary}}

In this work we have introduced a symmetry classification of translational symmetry broken particle-hole condensates, formulated in terms of representations of extended point groups. Extended point groups provide a natural way to study density wave formation at finite commensurate wave vector, as a generalization of point group classifications of density wave formation at nonzero angular momentum. In case of extended point group representations, ordering vector components take the role of angular momentum components. A prime example is the set of $M$-point wave vectors in systems with hexagonal symmetry, which are related by symmetry in a way analogous to $p$- or $d$-wave angular momentum form factors.

By way of example, we have applied the symmetry classification to two simple square lattice models. We then applied it to two hexagonal lattices: the triangular lattice and honeycomb lattices. In case of the hexagonal lattices, we find two common sets of $M$-point ordered density waves: a set of time-reversal even $s$ waves and a set of time-reversal odd $d$ waves, both owing their name to the form factors associated with each component of the set. We have considered the mean-field states corresponding to triple-$M$ ordering for both sets of density waves and found that the spectrum in one case ($s$ wave) is either gapped or exhibits an isolated touching of bands protected by symmetry, and is gapped in the other case ($d$ wave) with a Chern insulating mean-field ground state. 

These results only rely on symmetry and are therefore generally valid for lattices with hexagonal symmetry. We have verified this for the hexagonal kagome lattice. The general validity is confirmed by the derivation of the $M$-point nesting instabilities, presented in Sec.~\ref{sec:lowenergy}, which is based only on the low-energy $M$-point electrons. Given the nesting instabilities, we have developed a general theory of $M$-point electrons, with symmetry as the central ingredient, showing that two different gap structures can arise in case of triple-$M$ ordering. The gap structures can be expressed in terms of the diagonal Gell-Mann matrices, and describe a quadratic band touching in case of $s$-waves and a quadratic band crossing with time-reversal symmetry breaking gap in case of $d$ waves. These particular features of the low-energy theory of $M$-point electrons demonstrate the nontrivial nature of such theory, as it gives rise to electronic gapped or semi-metallic states with topological quantum numbers. 

In this work, we have paid special attention to hexagonal $M$-point orders associated with nesting instabilities. Interestingly, the $M$-point charge density waves (i.e., $s$ waves) have been found as ground states or leading instabilities of interacting electron models, in the context of triangular lattice~\cite{maharaj13}, honeycomb lattice~\cite{ye14}, and kagome lattice~\cite{kiesel13} models with extended Hubbard interactions, using Hartree-Fock and renormalization group methods. Similarly, a mean-field study of spinless electrons on the triangular lattice has reported spontaneous time-reversal symmetry breaking with $M$-point modulations~\cite{tieleman13}. Clearly, a full study of leading instabilities requires including the spin channel~\cite{martin08,makogon11,nandkishore12,kiesel12,wang12a,yu12,kiesel13,maharaj13,jiang14}. This study provides a stepping stone for a classification of $M$-point spin-triplet orders. A full symmetry classification of hexagonal spin-triplet orders will be presented elsewhere~\cite{venderbos}. 

The hexagonal lattice triple-$M$ $d$-wave state, shown in Figs.~\ref{fig:trif2} and~\ref{fig:honf2} and defined in Eqs.~\eqref{eq:f2tri} and~\eqref{eq:f2honey}, may be compared to the square lattice $d_{x^2-y^2}$ density wave. As is clear from the structure of these states, the $M$-point $d$-wave orders are the hexagonal counterpart of the square lattice $d$-wave (or staggered flux) state. However, whereas the square lattice has symmetry protected nodal degeneracies, as we showed in Section~\ref{sec:lowenergy}, the hexagonal $d$-waves, if realized as triple-$M$ order, lead to a gapped Chern insulating state. In case of the square lattice, the Chern insulating state requires superimposing density waves of two non-degenerate channels~\cite{laughlin98,kotetes08}. 

The hexagonal triple-$M$ $d$-wave state is a state with spontaneously generated current expectation values on lattice plaquettes, as is illustrated in Figs.~\ref{fig:trif2} and~\ref{fig:honf2}. In case of the square lattice $d$-density wave, it was pointed out that the statistical mechanics of such an arrangement of bond currents is governed by a six-vertex model~\cite{chakravarty02}. The six-vertex model is obtained by including, in addition to the two vertices of the homogeneously ordered $d$-density wave state, all other vertices that obey current conservation (i.e., absence of sources or sinks). The statistical mechanics of the six-vertex model is richer than that of a mean-field description, as current directions can be locally flipped, while keeping the amplitude of the bond order parameter fixed. The statistical mechanics of the hexagonal $d$-wave states is expected to be similar, with the important difference, however, that low-energy nodal quasiparticles are absent (the ordered ground state is insulating). Inspection of Fig.~\ref{fig:trif2} shows that on the triangular lattice, the statistical mechanics of $d$-wave state should be governed by a six-vertex model on a kagome lattice. Instead, Fig.~\ref{fig:honf2} shows that on the honeycomb lattice, the statistical mechanics should follow from an Ising model on a triangular lattice, since the current directions of all bonds of a current carrying hexagon have to be flipped due to current conservation. In this respect, the two symmetry-equivalent states are expected to lead to different behavior on the two lattices. 

A subject of increasing theoretical and experimental interest is the field of symmetry protected semimetals, with a specific focus on Dirac semimetals in three~\cite{young12} and two~\cite{young15} dimensions protected by nonsymmorphic crystal symmetry. Nonsymmorphic symmetries are crucial for protected nodal degeneracies at high-ssymmetry momenta, since, in general, only nonsymmorphic crystal symmetry groups allow for higher-dimensional representations at special points in the BZ. The classification formalism we present here, based on extended point group symmetry, is closely related to nonsymmorphic symmetry. The translations which are part of the extended point of the extended point group may be viewed as the fractional translations of the nonsymmorphic symmetry. As a result of this strong similarity, the extended point groups can be used as an alternative way to analyze degeneracy protection of semimetals when the effect of (fractional) translations is important. In Sec.~\ref{sec:lowenergy}, we showed how extended point group elements, specifically the composites of translations and ordinary point group elements, act nontrivially on electronic states at high-symmetry points [see, for instance, Eqs.~\eqref{eq:symvanhove} and~\eqref{eq:mrepvanhove}]. 
Therefore, extended point groups provide a direct route towards a systematic analysis of symmetry-protected semimetals when the crystal structure can be thought of as originating from broken translational symmetry.

\begin{acknowledgements}
I have benefited greatly from helpful and stimulating conversations and with L. Fu, C. Ortix, J. van Wezel, J. van den Brink, and M. Daghofer. I am particularly indebted to C. Ortix and J. van den Brink for a careful reading of the manuscript and thoughtful comments. This work was supported by NWO (The Netherlands).
\end{acknowledgements}

\appendix


\section{Definitions, conventions, and lattice symmetries\label{app:latsym}}

The atomic position, denoted $\vet{r}$, can be decomposed in terms of the Bravais lattice as $\vet{r} = \vet{x} + \vet{l}_{i}$, 
where $\vet{x}$ is a Bravais lattice vector and $\vet{l}_i$ denotes the position of the atom with respect to the unit cell vector $\vet{x}$. For a lattice with $N_{\text{sl}}$ different sublattices there are $N_{\text{sl}}$ distinct $\vet{l}$ vectors, i.e., $i=1,\ldots,N_{\text{sl}}$.  A lattice vector $\vet{x}$ can be written in terms of its generators as $\vet{x} = n_1 \vet{a}_1 + n_2 \vet{a}_2$, where $n_1,n_2$ are integers. 

The group of all spatial symmetries of the crystal lattice is given by the union of the point group $G$ and the group of translations $T$. Translations over a lattice vector are written as $T(\vet{x})$ and point group operations are denoted by $R$. 
For the purpose of this work we assume the equivalence of $D_n$ and $C_{nv}$, which is true for spinless electrons, and focus on $C_{nv}$. Any element of the space group can be written in terms of the four generators $T(\vet{a}_1)$, $T(\vet{a}_2)$, $C_n$ and $\sigma_v$, where $C_n$ is the $n$-fold rotation and $\sigma_v$ is a reflection $(x,y)\rightarrow (x,-y)$. Any element may then be specified by $T(\vet{a}_1)^{m_1}T(\vet{a}_1)^{m_2}C^{m_3}_n\sigma^{m_4}_v$ and point group operations $R$ can be written as $R=C^{m_1}_n\sigma^{m_2}_v$. 

The effect of a point group symmetry on an atomic position is represented as $R \vet{r}  = R\vet{x} + R\vet{l}_{i}$. This operation is a symmetry, hence $R \vet{r}$ is another atomic position, but possibly an inequivalent one. We have $R \vet{r}= \vet{r}'  = \vet{x}' + \vet{l}_{j}$. It is not necessarily true that $\vet{x}' =  R\vet{x} $, the difference must however be some lattice vector $\vet{t}_i$, $\vet{x}' =  R\vet{x} + \vet{t}_i$. $\vet{t}_i$ depends on the atom in the unit cell, hence the label $i$. It thus follows that $R \vet{r}  = R\vet{x} + \vet{t}_i+ \vet{l}_{j}$. 

\begin{figure}
\includegraphics[width=\columnwidth]{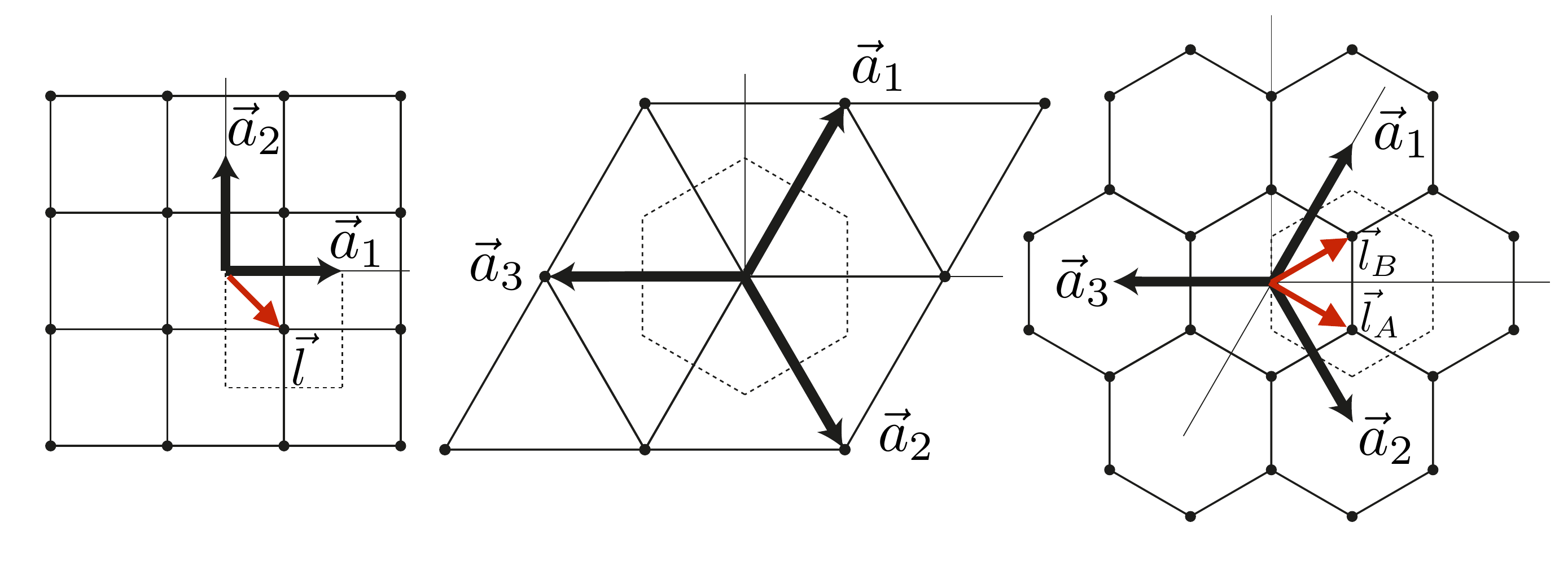}
\caption{\label{fig:lattices} Schematic illustration of the three lattices considered in this work. (left) the square lattice, (middle) triangular lattice, and (right) honeycomb lattice. Basis vectors $\vet{a}_i$ are represented as thick back arrows, the unit cell is given by dashed squares and hexagons. The red arrows indicate the vectors $\vet{l}_i$, i.e., the displacement of the atoms with respect to the origin.}
\end{figure}

The specific set of lattice vectors $\vet{a}_i$ and the displacement vectors $\vet{l}_i$ of the three lattices considered in this work are given in Table~\ref{tab:latvec} and shown in Fig.~\ref{fig:lattices}. Note that in case of the square lattice we choose the origin in the center of the square and not at an atomic site. 

\begin{table*}[t]
\centering
\begin{ruledtabular}
\begin{tabular}{cccc}
& Square &  Triangular &  Honeycomb\\ 
\hline
Lattice vectors  & $\vet{a}_1  = a(1,0)$  &  $\vet{a}_1  = a(1,\sqrt{3})/2$, $\vet{a}_2  = a(1,-\sqrt{3})/2$,   &  $\vet{a}_1  = a(1,\sqrt{3})/2$, $\vet{a}_2  = a(1,-\sqrt{3})/2$,  \\
   & $\vet{a}_2  = a (0,1)$ &  $\vet{a}_3 = -\vet{a}_1-\vet{a}_2$  &  $\vet{a}_3 = -\vet{a}_1-\vet{a}_2$  \\ [12pt]
 $\{ \vet{l}_i \}$ &$\vet{l}  =a (1,-1)/2$ & &  $\vet{l}_A=a( 1 , -1/\sqrt{3} )/2$,  \\
 &   & &  $\vet{l}_B=a( 1 , 1/\sqrt{3} )/2$\\ [12pt]
Nearest neighbors  &  &  &  $\vet{\delta}_1  = \vet{l}_B - \vet{l}_A$, $\vet{\delta}_2  =  \vet{\delta}_1 -\vet{a}_1$, $\vet{\delta}_3  = \vet{\delta}_1  +\vet{a}_2$  \\
\end{tabular}
\end{ruledtabular}
 \caption{Table of the lattice vectors of the square, triangular and honeycomb lattices. In case of square and honeycomb lattice the displacement vectors $\vet{l}_i$ are given, as well as the vectors $\vet{\delta}_i$ connecting nearest neighbors of the honeycomb lattice. }
\label{tab:latvec}
\end{table*}

To derive the transformation properties of the field operators and the Hamiltonian we define the annihilation (and creation) operators as $\anni{\sigma}{\vet{r}} = \anni{\sigma}{\vet{x} + \vet{l}_{i}} = \anni{i\sigma}{\vet{x} } \equiv \anniB{\vet{x}}$
(similarly for the creation operators). The index $\sigma$ labels spin and the label $i$ corresponds to the sublattice degrees of freedom. The Fourier transform of the field operators is given by $\anni{i\sigma}{\vet{k}}  = \sum_{\vet{x}}\anni{i\sigma}{\vet{x}}e^{-i\vet{k}\cdot \vet{x}}/\sqrt{N} $,
with $N$ the number of unit cells. With this definition we have $H(\vet{k}+\vet{G})= H(\vet{k})$. We define the operators $ \hat{U}_{R}$ and their Hermitian conjugates as acting on the field operators to implement the point group symmetry $R$. Then one has
\begin{gather}
\hat{U}_{R} \anni{\sigma}{\vet{r}} \hat{U}^\dagger_{R}  =\sum_{j\sigma'}  [D^\dagger_R]_{i\sigma j\sigma'} \anni{j\sigma'}{R\vet{x}} . \label{eq:symmatdef}
\end{gather}
Here, $D_R$ is a unitary matrix that acts in the space of internal orbital degrees of freedom ($\sigma$) and  in sublattice space ($i$). It may be viewed as a tensor product of matrices acting in each space separately.  
To deduce the transformation properties of the field operators in momentum space, we note that $\vet{x} + \vet{l}_{i} = R^{-1}(\vet{x}' + \vet{l}_j)$, where $\vet{x}' = R\vet{x}+\vet{t}_i$. Thus we get
\begin{gather} 
\hat{U}_{R} \anni{i\sigma}{\vet{k}} \hat{U}^\dagger_{R} = \sum_{j\sigma'} \sum_{\vet{x}} [D^\dagger_R]_{i\sigma j\sigma'} \anni{j\sigma'}{\vet{x}'} e^{-i\vet{k}\cdot \vet{x}} \nonumber \\
= \sum_{j\sigma'} \sum_{\vet{x}} [D^\dagger_R]_{i\sigma j\sigma'} \anni{j\sigma'}{\vet{x}'} e^{-i R\vet{k}\cdot R \vet{x} } \nonumber \\
= \sum_{j\sigma'} \sum_{\vet{x}} [D^\dagger_R]_{i\sigma j\sigma'} \anni{j\sigma'}{\vet{x}'} e^{-i R\vet{k}\cdot (\vet{x}' - \vet{t}_i)} \nonumber \\
 \label{eq:pgmom} = \sum_{j\sigma'}  [D^\dagger_R]_{i\sigma j\sigma'}  e^{i R\vet{k}\cdot \vet{t}_i}\anni{j\sigma'}{R\vet{k}} .
\end{gather}
For convenience, we define the new matrix $D_R(\vet{k})$ to include the $\vet{k}$ dependencies, i.e., we multiply $D^\dagger_R$ by $ \text{diag}(e^{iR\vet{k}\cdot \vet{t}_1},\ldots,e^{iR\vet{k}\cdot \vet{t}_{N_{\text{sl}}}}) $ (from the left). We then obtain $\hat{U}_{R} \anniB{\vet{k}} \hat{U}^\dagger_{R} = D^\dagger_R(\vet{k}) \anniB{R\vet{k}}$.

The translationally invariant Hamiltonian is generically written as
\begin{gather*}
\hat{H}  = \sum_{ij\sigma\sigma'}\sum_{\vet{k}} \crea{i\sigma}{\vet{k}} H_{i\sigma j\sigma'}(\vet{k}) \anni{j\sigma'}{\vet{k}}. 
\end{gather*}
Under the symmetry operation the Hamiltonian transforms as 
\begin{gather*}
\hat{U}_{R} \hat{H} \hat{U}^\dagger_{R} =  \sum_{ij\sigma\sigma}\sum_{\vet{k}} (D_R HD^\dagger_R)_{i\sigma j\sigma'}(\vet{k}) \crea{i\sigma}{R\vet{k}}\anni{j\sigma'}{R\vet{k}},
\end{gather*}
and since this is a symmetry we must have $\hat{U}_{R} \hat{H} \hat{U}^\dagger_{R} = \hat{H}$, which implies
\begin{gather}
D_R(\vet{k}) H(\vet{k}) D^\dagger_R(\vet{k})  = H(R\vet{k}).
\end{gather}

The composition of two point group symmetries $R_1$ and $R_2$ yields another element $R_3=R_2R_1$. The transformation properties in case of a product of symmetries on the field operator is [using Eq.~\eqref{eq:pgmom}],
\begin{gather*} 
\hat{U}_{R_3} \anniB{\vet{k}} \hat{U}^\dagger_{R_3} = \hat{U}_{R_2} \hat{U}_{R_1} \anniB{\vet{k}} \hat{U}^\dagger_{R_1} \hat{U}^\dagger_{R_2} \\
= D^\dagger_{R_1}(\vet{k})D^\dagger_{R_2}(R_1\vet{k}) \anniB{R_2R_1\vet{k}}.
\end{gather*}
If and only if $R\vet{k}^* = \vet{k}^*$ for some $\vet{k}^*$ in the BZ, and \emph{all} $R$ of the point group, do the $ D_{R}$ form a representation of the group, i.e., $ D_{R_3}=  D_{R_2} D_{R_1}$. If a proper subgroup of the point group leaves a certain $\vet{k}^*$ invariant, then the $D_{R}$ will form a representation of that subgroup.

The Fourier transform reflects the requirement $H(\vet{k}+\vet{G})= H(\vet{k})$. A common alternative definition used in the context of tight-binding models is (suppressing spin) $\anni{i}{\vet{k}}=\sum_{\vet{x}}\anni{i}{\vet{x}}e^{-i\vet{k}\cdot (\vet{x}+\vet{l}_i)} /\sqrt{N} = e^{-i\vet{k}\cdot \vet{l}_i } \sum_{\vet{x}}\anni{i}{\vet{x}}e^{-i\vet{k}\cdot \vet{x}} /\sqrt{N}  $.
The two conventions are related by a gauge transformation of the form $A(\vet{k}) = \text{diag}(e^{i\vet{k}\cdot \vet{l}_1 },\ldots ,e^{i\vet{k}\cdot \vet{l}_{N_{\text{sl}}} })$. 
Hence, the momentum-dependent Hamiltonian in the tight-binding basis is written in terms of the definition above as $H'(\vet{k}) = A^\dagger(\vet{k})  H(\vet{k})A(\vet{k})$, and we have that $H'(\vet{k}+\vet{G}) \neq H'(\vet{k})$.

It is important to be aware of this when studying invariant $\vet{k}$ points, i.e., momenta for which $R\vet{k}^* = \vet{k}^*\, \text{mod}\, \vet{G}$. In that case one often needs precisely $H(\vet{k}+\vet{G}) = H(\vet{k})$, so that $H(R\vet{k}^*)  = H(\vet{k}^*\, \text{mod}\, \vet{G}) = H(\vet{k}^*)$.

\subsection{Lattice symmetries in the reduced or folded zone\label{ssec:symtsb}}

Translational symmetry breaking due to finite wave vector modulations couples electrons with relative momentum equal to the wave-vector. To describe translational symmetry breaking, the dimension of the Hamiltonian (i.e., equivalently, the unit cell) is enlarged, since $\anni{}{\vet{k}}$ and $\anni{}{\vet{k}+\vet{Q}}$ may be coupled. Here, we present the effective mean-field description and the action of lattice symmetries for the case of a set of wave vectors $\vet{Q}_\mu$ satisfying $2 \vet{Q}_\mu=0$ and $\vet{Q}_1+\vet{Q}_2+\vet{Q}_3=0$, which is relevant to cases discussed in this paper. 

We write the mean-field Hamiltonian defined over the reduced BZ as $\hat{H} = \sum_{\vet{k}\in \text{rbz}}  \hat{\chi}^\dagger(\vet{k}) \mathcal{H}(\vet{k}) \hat{\chi}(\vet{k}) $
where the mean-field annihilation operator $\hat{\chi}$ is given by
\begin{gather} \label{eq:mfspinor}
\hat{\chi}(\vet{k}) = \hat{\chi}_{\mu j}(\vet{k}) =  \begin{pmatrix} \hat{\chi}_{0j}(\vet{k}) \\ \hat{\chi}_{1j}(\vet{k}) \\ \hat{\chi}_{2j}(\vet{k}) \\ \hat{\chi}_{3j}(\vet{k})  \end{pmatrix} = \begin{pmatrix} \anni{j}{\vet{k}} \\ \anni{j}{\vet{k}+\vet{Q}_1} \\ \anni{j}{\vet{k}+\vet{Q}_2} \\ \anni{j}{\vet{k}+\vet{Q}_3} \end{pmatrix}.
\end{gather}

The action of point group operations on $\hat{\chi}(\vet{k}) $ is derived as (suppressing the internal index $\sigma$, and the explicit dependence of unitary transformations on $R$)
\begin{widetext}
\begin{gather} 
\hat{U} \hat{\chi}_{\mu i}(\vet{k}) \hat{U}^\dagger = \hat{U} \anni{i}{\vet{k}+\vet{Q}_\mu}  \hat{U}^\dagger = 
 D^\dagger_{ij}(\vet{k}+\vet{Q}_\mu) \anni{j}{R\vet{k}+R\vet{Q}_\mu}  = 
 D^\dagger_{ij}(\vet{k}+\vet{Q}_\mu) V^\dagger_{\mu\nu} \anni{j}{R\vet{k}+\vet{Q}_\nu}  \nonumber \\
=D^\dagger_{ij}(\vet{k}+\vet{Q}_\mu) V^\dagger_{\mu\nu}  \hat{\chi}_{\nu j}(R\vet{k})=  
  \label{eq:tsb}\begin{pmatrix}  D^\dagger (\vet{k})  & & \\  & \ddots & \\  & & D^\dagger(\vet{k}+\vet{Q}_3) \end{pmatrix}_{\mu\nu} V^\dagger_{\nu\eta}\hat{\chi}_{\eta}(R\vet{k}) .
\end{gather}
\end{widetext}
In this expression, $V$ is a matrix acting on the momentum components $\mu$ of $\hat{\chi}_\mu$. The point group operation $R$ generally permutes the momenta $\vet{Q}_\mu$ and $V$ implements this permutation. 

Based on the transformation of $\hat{\chi}$ we obtain an expression for the symmetry condition on the mean-field Hamiltonian. In analogy with Eq.~\eqref{eq:symmatdef}, we define the matrix $\mathcal{D}(\vet{k})$ as $\hat{U} \hat{\chi}(\vet{k}) \hat{U}^\dagger \equiv \mathcal{D}^\dagger(\vet{k})\hat{\chi}(R\vet{k}) $ (obviously $\mathcal{D} = \mathcal{D}_R$, which we suppress for convenience). Invariance of the Hamiltonian under $R$ is then expressed as
\begin{gather} \label{eq:hamtranssb}
\mathcal{D}(\vet{k}) \mathcal{H}(\vet{k}) \mathcal{D}^\dagger(\vet{k})  = \mathcal{H}(R\vet{k}) 
\end{gather}
Here, $\vet{k}$ is restricted to the reduced BZ (RBZ). Care must be taken when analyzing invariant $\vet{k}$-points in the reduced BZ, as these are generally not invariant points in the original BZ. To demonstrate this, let us assume that $R$ leaves $\vet{k}^*$ invariant in the RBZ. We then have $\hat{\chi}_{\mu }(R\vet{k}^*) = \hat{\chi}_{\mu }(\vet{k}^*+\vet{G}_{\text{rbz}})$, where $\vet{G}_{\text{rbz}}$ is a reciprocal lattice vector of the RBZ. For $\hat{\chi}_{\mu}(\vet{k})$ defined in Eq.~\eqref{eq:mfspinor}, $\vet{G}_{\text{rbz}}$ is simply one of the $\vet{Q}_\mu$ since these define the reciprocal lattice vectors of the RBZ. The vectors $\vet{Q}_\mu$ form a group under addition and therefore $\vet{k}^*+\vet{G}_{\text{rbz}}$ is equal to a permutation of components of $\hat{\chi}_{\mu }$, i.e., $\hat{\chi}_{\mu }(\vet{k}^*+\vet{G}_{\text{rbz}}) = W^{\dagger}_{\mu\nu}\hat{\chi}_{\nu }(\vet{k}^*) $. The matrix $W_{\mu\nu}$ implements the equivalence of momenta in the RBZ. In particular, this implies for a symmetry $R$ at $\vet{k}^*$
\begin{gather} \label{eq:hamequiv}
W \mathcal{H}(R\vet{k}^*) W^\dagger = W\mathcal{H}(\vet{k}^* +\vet{G}_{\text{rbz}}) W^\dagger =  \mathcal{H}(\vet{k}^* ) 
\end{gather}
and, consequently,  
\begin{gather} \label{eq:hamcom}
W \mathcal{D}(\vet{k}^*) \mathcal{H}(\vet{k}^*) \mathcal{D}^\dagger(\vet{k}^*) W^\dagger=  \mathcal{H}(\vet{k}^* ).
\end{gather}

We note in passing that it is straightforward to change to a gauge for which $\mathcal{H}(\vet{k} +\vet{G}_{\text{rbz}}) =  \mathcal{H}(\vet{k} ) $, where $ \mathcal{H}(\vet{k}) $ is the mean-field Hamiltonian. One may choose $\vet{Q}_1$ and $\vet{Q}_2$ as generators of the reciprocal lattice, for which we have $\hat{\chi}(\vet{k}+\vet{Q}_{1}) = W^\dagger_{1}\hat{\chi}(\vet{k}) $ and $\hat{\chi}(\vet{k}+\vet{Q}_{2}) = W^\dagger_{2}\hat{\chi}(\vet{k}) $. Clearly $\hat{\chi}(\vet{k}+\vet{Q}_{1}+\vet{Q}_2) = W^\dagger_{1}W^\dagger_{2}\hat{\chi}(\vet{k})=W^\dagger_{2}W^\dagger_{1}\hat{\chi}(\vet{k}) $, implying that $W_{1}$ and $W_{2}$ commute and are simultaneously diagonalizable. For $2\vet{Q}_{1}=2\vet{Q}_{2}=0$ we have in addition $(W_{1})^2=(W_{2})^2=1$ mandating the eigenvalues to be $e^{i \phi_\mu}$ with $\phi_\mu=0,\pi$. One now sets $\phi_\mu = \vet{k}\cdot \vet{a}_{1,2}$ so as to match correct value for $\vet{Q}_1\cdot \vet{a}_{1,2} = 0,\pi$ and $\vet{Q}_2\cdot \vet{a}_{1,2}=0,\pi$ simultaneously. This then defines the gauge transformation needed to compensate the eigenvalues of $U^{\text{eq}}_{1}$ and $U^{\text{eq}}_{2}$.


\section{General constraints and properties of condensate functions~\label{ssec:compat}}

Condensation at finite commensurate ordering vector imposes constraints on the condensate functions~\cite{nayak00}. Here we briefly discuss these constraints for the two sets of commensurate ordering vectors. The first set consists of the triple $\vet{Q}_\mu$ ($\mu=1,2,3$) with the properties $2\vet{Q}_\mu = 0$ and $\pm \vet{Q}_1 \pm \vet{Q}_2 \pm \vet{Q}_3 = 0$. It describes the vectors $(\vet{Q},\vet{X},\vet{Y})$ in case of the square lattice and the triple $(\vet{M}_1,\vet{M}_2,\vet{M}_3)$ in case of hexagonal lattices. In terms of the general $\vet{Q}_\mu$, the condensate functions are defined as
\begin{gather}
\order{\crea{\sigma}{\vet{k}+\vet{Q}_{\mu} }\anni{\sigma' }{\vet{k}}} = \Delta_\mu(\vet{k})\delta_{\sigma\sigma'},
\end{gather}
where for the moment we neglect sublattice degrees of freedom. Using commensurability, $2\vet{Q}_\mu = 0$, it is straightforward to find the relation
\begin{gather} \label{eq:compat}
\frac{\Delta_\mu(\vet{k}+\vet{Q}_\mu)}{\Delta_\mu^*(\vet{k})} = 1.
\end{gather}
The condensate function $\Delta_\mu(\vet{k}) $ can be written as $\Delta_\mu(\vet{k}) = \Delta_\mu \lambda_\mu(\vet{k})$ with $\Delta_\mu $ a complex order parameter and $\lambda_\mu(\vet{k})$ an orbital function transforming as an irreducible representation of the group of the wave vector $\vet{Q}_\mu$. Then, Eq.~\eqref{eq:compat} takes the form $\lambda_\mu(\vet{k}+\vet{Q}_\mu) / \lambda^*_\mu(\vet{k})  = \Delta^*_\mu / \Delta_\mu$. For $\vet{Q}_\mu$ satisfying $2\vet{Q}_\mu=0 $, one generally has $\lambda_\mu(\vet{k}+\vet{Q}_\mu) = \pm \lambda_\mu(\vet{k})$ and together with the property $\lambda^*_\mu(\vet{k}) = \lambda_\mu(\vet{k})  $, which is true for the square and triangular lattices, one obtains $\Delta^*_\mu  = \pm  \Delta_\mu $. For commensurate ordering $\Delta_\mu$ is either real or imaginary which implies the absence of a phase degree of freedom and the Goldstone modes associated with it. 

The property $\pm \vet{Q}_1 \pm \vet{Q}_2 \pm \vet{Q}_3 = 0$ can be used to establish that 
\begin{eqnarray}
\Delta_1 (\vet{k}) &=&  \order{\crea{\sigma}{\vet{k}+\vet{Q}_1 }\anni{\sigma' }{\vet{k}}}  \nonumber \\
&= &\order{\crea{\sigma}{\vet{k}+\vet{Q}_2+\vet{Q}_3 }\anni{\sigma' }{\vet{k}+\vet{Q}_2+\vet{Q}_2 }}   \nonumber \\
 &= &\order{\crea{\sigma}{\vet{k}+\vet{Q}_2+\vet{Q}_3 }\anni{\sigma' }{\vet{k}+\vet{Q}_3+\vet{Q}_3 }} 
\end{eqnarray}
from which it easily follows that
 \begin{eqnarray}
\Delta_1 (\vet{k}-\vet{Q}_2) & = &  \order{\crea{\sigma}{\vet{k}+\vet{Q}_3 }\anni{\sigma' }{\vet{k}+\vet{Q}_2 }}  \nonumber \\
=\Delta_1^* (\vet{k}-\vet{Q}_3)  & = & \order{\crea{\sigma}{\vet{k}+\vet{Q}_2 }\anni{\sigma'}{\vet{k}+\vet{Q}_3 }}^* 
\end{eqnarray}
Similar relations hold for the other combinations of ordering momenta. 

Time-reversal symmetry $\Theta$ acts as $\Delta_\mu(\vet{k}) \rightarrow \Delta^*_\mu(-\vet{k})$. Combining $\Delta_\mu(\vet{k}) = \Delta_\mu \lambda_\mu(\vet{k})$ and $\Delta^*_\mu  = \pm  \Delta_\mu$ one has $\Delta^*_\mu(-\vet{k}) = \pm \Delta_\mu \lambda_\mu(-\vet{k})$. Thus, the parity of $\lambda_\mu(\vet{k})$ determines whether the condensate function is time-reversal even or odd. 

In the presence of a sublattice degree of freedom, which is the case for the honeycomb lattice, the condensate function is a matrix in sublattice space,   
\begin{gather}
\order{\crea{i \sigma }{\vet{k}+\vet{Q}_\mu} \anni{j \sigma' }{\vet{k}}} = [\hat{\Delta}_{\mu}(\vet{k})]_{ij}.
\end{gather}
Using the properties of $\vet{Q}_\mu$ one finds
\begin{gather} \label{eq:hermitian}
\order{\crea{i \sigma }{\vet{k}} \anni{j \sigma' }{\vet{k}+\vet{Q}_\mu}} = [\hat{\Delta}_{\mu}(\vet{k})]^\dagger_{ij}= [\hat{\Delta}_{\mu}(\vet{k}+\vet{Q}_\mu)]_{ij}.
\end{gather}
In general, this does not constrain the off-diagonal elements to have purely real or purely imaginary $\hat{\Delta}_{\mu}$. However, since time-reversal symmetry $\Theta$ acts as $[\hat{\Delta}_{\mu}(\vet{k})]_{ij} \rightarrow  [\hat{\Delta}_{\mu}(-\vet{k})]^*_{ij}$, the real and imaginary parts of the off-diagonal elements are even and odd under time reversal, depending on whether the orbital functions are even or odd. 

The second set of commensurate ordering vectors is given by the hexagonal symmetry $K$ points, i.e., $(\vet{K}_+,\vet{K}_-)$. Their commensurability is expressed as $3\vet{K}_+=3\vet{K}_-=0$ and $\vet{K}_+=-\vet{K}_-$. Writing a general condensate matrix as 
\begin{gather}
\order{\crea{i \sigma }{\vet{k}+\vet{K}_\pm} \anni{j \sigma' }{\vet{k}}} = [\hat{\Delta}_{\pm}(\vet{k})]_{ij},
\end{gather}
and using the commensurability relations one obtains 
\begin{gather}
 [\hat{\Delta}_{\pm}(\vet{k})]^\dagger_{ij} = [\hat{\Delta}_{\pm}(\vet{k}+\vet{K}_\mp)]_{ij}.
\end{gather}
In addition, one finds that 
\begin{gather}
\order{\crea{i \sigma }{\vet{k}+\vet{K}_\mp} \anni{j \sigma' }{\vet{k}+\vet{K}_\pm}} = [\hat{\Delta}_{\pm}(\vet{k}+\vet{K}_\pm)]_{ij} 
\end{gather}
When there is no sublattice degree of freedom (and for the diagonal elements $[\hat{\Delta}_{\pm}]_{ii}$), the functions $\Delta_\pm(\vet{k})$ are related by $\Delta_-(\vet{k}) = \Delta^*_+(\vet{k}+\vet{K}_-) $.  Time-reversal symmetry $\Theta$ acts as $\Delta_+(\vet{k}) \rightarrow \Delta^*_-(-\vet{k}) $. For the simplest case of $\Delta_\pm(\vet{k}) = \Delta_\pm $, i.e., no momentum dependence, we have $\Delta_- = \Delta^*_+$, which automatically respects time-reversal. Note that $\Delta_\pm$ can be complex. 

\section{Degeneracies at the $M$ and $M'$ points\label{app:mpoint}}

In Sec.~\ref{sec:lowenergy}, we argued that degeneracies at the $M'$ points of the folded Brillouin zone in general are not protected unless translational symmetry is preserved. Here, we show this by considering the action of translations and $C_{2v}$ symmetry on $\hat{\Phi}_{\text{M}'}$ given by 
\begin{gather}
\hat{\Phi}_{\text{M}'} = \begin{pmatrix} \anni{}{\vet{M}'}  \\ \anni{}{\vet{M}'+\vet{M}_1} \\ \anni{ }{\vet{M}'+\vet{M}_2}   \\ \anni{ }{\vet{M}'+\vet{M}_3}   \end{pmatrix},
\end{gather}
where $\vet{M}' = \vet{M}_2/2$ by way of example. The action of, for instance, the translation $T(\vet{a}_1)$ is expressed as 
\begin{gather}
T(\vet{a}_1) \; \sim \;  \begin{pmatrix} \tau^3 &\\ & -\tau^3 \end{pmatrix} \equiv \nu^3\tau^3,
\end{gather}
defined in terms of Pauli matrices $\tau^i$ and $\nu^i$. For the inversion $C_2$ one finds
\begin{gather}
C_2 \; \sim \;  \begin{pmatrix}  &1 \\ 1&\end{pmatrix} \equiv \nu^1.
\end{gather}
Note that we have used $-\vet{M}' = \vet{M}' + \vet{M}_2$. Symmetry dictates that the Hamiltonian at $M'$ must commute with these elements. Considering only the translation and the inversion we see that the only allowed term is $\tau^3$. Hence, eigenvalues of the Hamiltonian at $\vet{M}' $ necessarily come in pairs. Breaking translational symmetry reduces the symmetry to $C_{2v}$, which cannot protect degeneracies. 

In a similar manner as the triangular lattice, we can consider degeneracies of the honeycomb lattice at the $M$ and $M'$ points. The analysis and the results are the same, the only difference is the sublattice degree of freedom. The electron operator $\hat{\Phi}_{\Gamma}$ at $\Gamma$ of the folded zone is
\begin{gather} \label{eq:honMstate}
\hat{\Phi}_{\Gamma} =  \begin{pmatrix} \hat{\chi}_{1j} \\ \hat{\chi}_{2j}   \\ \hat{\chi}_{3j}  \end{pmatrix}= \begin{pmatrix}  \anni{j }{\vet{M}_1} \\ \anni{j }{\vet{M}_2}   \\ \anni{j }{\vet{M}_3}   \end{pmatrix},
\end{gather}
where $j$ is the sublattice index. The action of generators of $C'''_{6v}$ is derived using formula~\eqref{eq:tsb}. For instance, for the sixfold rotation $C_6$ one finds the matrix
\begin{gather} \label{eq:MrepC}
C_6 \quad \hat{\Phi}_{\Gamma}  \;  : \; \rightarrow \; \begin{pmatrix}   & \tau^3\tau^1 &  \\ &  & \tau^3\tau^1 \\ \tau^1 &&   \end{pmatrix}\hat{\Phi}_{\Gamma} ,
\end{gather}
where now $\tau^3=\pm 1$ is the sublattice index. The matrix representation obtained in this way is six dimensional and can be decomposed into irreducible representations of $C'''_{6v}$, and one finds $F_1+F_4$. As a consequence, there are two protected three-fold degeneracies in the presence of full $C'''_{6v}$ symmetry. The two sets of van Hove operators $\hat{\Phi}_{F_1} $ and $\hat{\Phi}_{F_4} $ are given by
\begin{gather}
\hat{\Phi}_{F_1} =  \begin{pmatrix} \hat{\chi}_{1A} - \hat{\chi}_{1B}   \\ \hat{\chi}_{2A} + \hat{\chi}_{2B}     \\   -\hat{\chi}_{3A} + \hat{\chi}_{3B}   \end{pmatrix}, \; \hat{\Phi}_{F_4} =  \begin{pmatrix} \hat{\chi}_{1A} + \hat{\chi}_{1B}   \\ \hat{\chi}_{2A} - \hat{\chi}_{2B}     \\   -\hat{\chi}_{3A} - \hat{\chi}_{3B}   \end{pmatrix}.
\end{gather}
The analysis presented in Sec.~\ref{sec:lowenergy} directly applies to these sets of van Hove operators. 


\section{$M$-point representation of hexagonal symmetry\label{app:malgebra}}

Here, we introduce a representation of the hexagonal symmetry group associated with the three inequivalent $M$ points. The representation is defined by the action of elements of the symmetry group on the vector $\vet{v}=\vet{v}(\vet{x})$:
\begin{gather} \label{eq:vvector}
\vet{v}(\vet{x}) = \begin{pmatrix} \cos \vet{M}_1\cdot\vet{x} \\ \cos \vet{M}_2\cdot\vet{x}  \\  \cos \vet{M}_3\cdot\vet{x}  \end{pmatrix}.
\end{gather} 
The components of $\vet{v}$ are the linearly independent functions of $\vet{x}$, with modulations set by $\vet{M}_\mu$. Note that $\sin \vet{M}_\mu\cdot\vet{x} = 0$ since $\vet{M}_\mu\cdot\vet{x} = 0,\pm \pi$ for lattice vectors $\vet{x}$. In addition, due to $2\vet{M}_\mu = 0$ one has $\cos \vet{M}_\mu\cdot\vet{x} = e^{i \vet{M}_\mu\cdot\vet{x} }$. 

The elementary translations $T(\vet{a}_i)$ are represented by the matrices $G_i$ defined through the equation
\begin{gather} \label{eq:defG}
\vet{v}(\vet{x}+\vet{a}_i) \equiv G_{i}\vet{v}(\vet{x}) , \quad i=1,2,3.
\end{gather}
Explicitly, $G_1$ and $G_2$ are given by
\begin{gather} 
G_{1}= \begin{pmatrix}-1 && \\ &-1& \\ &&1 \end{pmatrix}, \;  G_{2} = \begin{pmatrix}1 && \\ &-1& \\ &&-1 \end{pmatrix}.
\end{gather}
These matrices have the property $G^2_i = 1$, they mutually commute, and multiplication of two of them gives the third, i.e., $G_1G_2=G_3$, etc. This is direct consequence of $M$-point modulations and the algebra of translations $T(\vet{a}_i)$ in the group $C'''_{6v}$.

Rotations and reflections can be expressed in terms of the generators $C_6$ and $\sigma_v$. The action of $C_6$, defined as $\vet{v}'(\vet{x}) = \vet{v}(C^{-1}_6\vet{x})$, is represented by the matrix $X$ and given by
\begin{gather} \label{eq:defX}
\vet{v}(C^{-1}_6\vet{x}) = X \vet{v}(\vet{x}), \quad  X = \begin{pmatrix}  0 & 1 & 0 \\ 0 & 0 & 1 \\1 & 0 & 0 \end{pmatrix}.
\end{gather}
Note that $X$ has the property $X^{3}=1$ and thus $X^{-1} = X^2$. In addition, one has $X^{-1} =  X^T$, where $X^T$ is the transpose. It thus follows that $\vet{v}(C^{-1}_3\vet{x}) = X^2 \vet{v}(\vet{x}) =  X^T \vet{v}(\vet{x})$. For the reflection $\sigma_v$, we define the matrix $Y$ as
\begin{gather}  \label{eq:defY}
\vet{v}(\sigma^{-1}_v \vet{x}) = Y\vet{v}(\vet{x}), \quad  Y = \begin{pmatrix}  0 & 0 & 1 \\ 0 & 1 & 0 \\ 1 & 0 & 0 \end{pmatrix}.
\end{gather}
All rotations and reflections can be represented by a product of powers of $X$ and $Y$, i.e., $X^mY^n$. An arbitrary string of $X$ and $Y$ matrices can be brought into this form using $(XY)^2=1$, which is equivalent to $XY=YX^T$. 

As a representation of the group $C'''_{6v}$, the representation defined by $G_j$, $X$, and $Y$ is irreducible, and computing characters shows it is equal to $F_1$. As a representation of $C_{6v}$, the representation defined by $X$ and $Y$ is reducible and the decomposition is $A_1 + E_2$. 

We can interpret the matrices $G_i$, $X$, and $Y$ as $O(3)$ rotation matrices, showing that these matrices define an embedding of $C'''_{6v}$ in $O(3)$. This does not, however, define an invertible map between $C'''_{6v}$ and $O(3)$, since $C_2 = C^3_6$ is mapped to identity through $X^3=1$. This may be remedied by redefining $-X$ as the $O(3)$ matrix corresponding to the generator $C_6$. The two-fold rotation $C_2 \in C'''_{6v}$ is then mapped to the inversion $P \in O(3)$. 
Commutators of the elements $G_i$, $X$, and $Y$ can be obtained by direct computation using~\eqref{eq:defG}--\eqref{eq:defY}.

The hexagonal $M$-point representation in terms of $O(3)$ matrices can be defined alternatively by considering the effect of hexagonal symmetry on electron operators at the $M$ points. Defining $\hat{\Phi} $ as the $M$-point electron operator [see Eq.~\eqref{eq:vanhovehexa} of main text], i.e.,
\begin{gather}
\hat{\Phi} =  \begin{pmatrix} \anni{ }{\vet{M}_1} \\ \anni{}{\vet{M}_2}   \\ \anni{ }{\vet{M}_3}   \end{pmatrix},
\end{gather}
we can evaluate the action of elements of $C'''_{6v}$. The action of the three generators is given by 
\begin{align} 
T(\vet{a}_1) \; : \;  \hat{\Phi} &\rightarrow \; G_1 \hat{\Phi}, \nonumber \\
C_6 \; : \;  \hat{\Phi} &\rightarrow \; X \hat{\Phi} \nonumber \\
\sigma_v \; : \;  \hat{\Phi} &\rightarrow \; Y \hat{\Phi},  \label{eq:vanhoveapp}
\end{align}
where $G_1$, $X$, and $Y$ are defined as in~\eqref{eq:defG}--\eqref{eq:defY}, showing that $\hat{\Phi} $ defines the same representation. 

With the help of this representation all fermion bilinears $\hat{\Lambda}$ given by
\begin{gather}
\hat{\Lambda}= \hat{\Phi}^\dagger_\mu \Lambda_{\mu\nu}  \hat{\Phi}_\nu ,
\end{gather}
can be classified in terms of symmetry. Here $\Lambda$ is an Hermitian matric. The space of these $M$-point Hermitian matrices is spanned by the Gell-Mann matrices, the generators of SU(3). We group them in three sets defined by $\vet{\Lambda}_a$, $\vet{\Lambda}_b$, and $\vet{\Lambda}_c$. 
Evaluating the transformation properties under $C'''_{6v}$ one finds that $\vet{\Lambda}_a$ transforms as $F_1$ and $\vet{\Lambda}_b$ as $F_2$.


\section{Real space construction of $M$-point order\label{app:mreal}}

In this appendix, we show how explicit expressions of density waves can be systematically derived in real space, with the help of the $M$-point representation defined and discussed in detail in Appendix~\ref{app:malgebra}. We first consider the triangular lattice and then the honeycomb lattice. 

\subsection{Triangular lattice}

First, we seek to obtain the $M$-point site ordered states of $F_1$ symmetry, Eq.~\eqref{eq:trisitem}. The starting point is the real space condensate expression
\begin{gather}
\order{\crea{ \sigma }{\vet{x}} \anni{ \sigma' }{\vet{y}}} = \Delta \;  \vet{w}\cdot \vet{v}(\vet{x}) \;\delta_{\vet{x},\vet{y}}\delta_{\sigma\sigma'},
\end{gather} 
The inner product $\vet{w}\cdot \vet{v}(\vet{x})$ is a concise way of writing a general linear combination of the modulation functions $\cos \vet{M}_\mu\cdot \vet{x}$ [see Eq.~\eqref{eq:vvector}]. In general, it turns out to be convenient to directly derive the triple-$M$ ordered state, in this case with $A_1$ symmetry. Elements of $C_{6v}$ act on $\vet{w}$ via $\vet{v}(\vet{x})$. For invariance under the three-fold rotation and the reflection $\sigma_v$, we find the conditions $X\vet{w}=\vet{w} $ and $Y\vet{w}=\vet{w}$, respectively. This fixes $\vet{w}$ to be $\vet{w}\sim (1,1,1)$. Taking the Fourier transform gives Eq.~\eqref{eq:f1tri}.

Next, we take the flux ordered states with $F_2$ symmetry, Eq.~\eqref{eq:trifluxm}. In that case, the starting point is the real space condensate for bonds in the $\vet{a}_1$ direction
\begin{gather} \label{eq:rstriflux}
\order{\crea{ \sigma }{\vet{x}} \anni{ \sigma' }{\vet{y}}} = \Delta \;  i \vet{w}\cdot \vet{v}(\vet{x}) \;(\delta_{\vet{x}+\vet{a}_1,\vet{y}}-\delta_{\vet{x},\vet{y}+\vet{a}_1}) \delta_{\sigma\sigma'}.
\end{gather} 
Expressions for bonds along the $\vet{a}_2$ and $\vet{a}_3$ directions are obtained using the three-fold rotation, and given by $X^T\vet{w}$ and $X\vet{w}$, respectively. The triple-$M$ ordered state has $A_2$ symmetry. Invariance under inversion gives the relation $G_1 \vet{w} = -\vet{w}$. The solution is $\vet{w}\sim (1,\pm 1,0)$. The reflection $\sigma_v$ fully determines $\vet{w}$ leading to $YX\vet{w}= -\vet{w}$ and $\vet{w}\sim (1,- 1,0)$. Fourier transforming gives Eq.~\eqref{eq:f2tri}.

\subsection{Honeycomb}

We seek to derive honeycomb lattice site and bond order with $F_1$ symmetry and flux order with $F_2$ symmetry. Site order can be generally expressed in terms of the vectors $ \vet{w}_i$ ($i=A,B$): 
\begin{gather}
\order{\crea{i \sigma }{\vet{x}} \anni{j \sigma' }{\vet{y}}} = \Delta \;  \vet{w}_i\cdot \vet{v}(\vet{x}) \;\delta_{\vet{x},\vet{y}}\delta_{ij}\delta_{\sigma\sigma'}.
\end{gather} 
The vectors $ \vet{w}_i$ contain the order parameter components for each sublattice, to be determined by evaluating symmetry constraints. A convenient route is to start from the threefold rotation $C_3$. Evaluating the effect of $C_3$ and $C^2_3$ on $ \vet{w}_A$ yields $G_2X \vet{w}_A$ and $(G_2X)^2\vet{w}_A$, respectively. We therefore consider the matrix operator $P = (1+G_2X^T+ G_2X^TG_2X^T)/3$, which is invariant under $C_3$. We find that $P^2 = P$, meaning that $P$ is a projector having eigenvalues $0$ and $1$. We look for the eigenvector corresponding to eigenvalue $1$ and find $\vet{w}_A= (-1,-1,1)$ (the null space of $P$ is two dimensional). To find $ \vet{w}_B$, we evaluate a symmetry that exchanges the sublattices and relates $\vet{w}_B$ and $\vet{w}_A$. We may take the inversion $C_2$, which leads to $G_3 \vet{w}_A  = \pm \vet{w}_B$. The positive solution corresponds to $F_1$ and fixes $\vet{w}_B = ( 1,-1,-1)$, yielding the vectors quoted prior to Eq.~\eqref{eq:f1honey}. [Note that the $F_4$ representation of Eq.~\eqref{eq:hexasitem} is obtained by inverting the sign of $\vet{w}_B$.]

Bond order condensate functions can be parametrized by $\vet{w}_i$ with $i=1,2,3$, 
\begin{eqnarray}
\order{\crea{A \sigma }{\vet{x}} \anni{B \sigma' }{\vet{y}}} &=& \Delta \;  \vet{w}_1\cdot \vet{v}(\vet{x}) \;\delta_{\vet{x},\vet{y}}\delta_{\sigma\sigma'} \nonumber \\
\order{\crea{A \sigma }{\vet{x}} \anni{B \sigma' }{\vet{y}}} &=& \Delta \;  \vet{w}_2\cdot \vet{v}(\vet{x}) \;\delta_{\vet{x}-\vet{a}_1,\vet{y}}\delta_{\sigma\sigma'} \nonumber \\
\order{\crea{A \sigma }{\vet{x}} \anni{B \sigma' }{\vet{y}}} &=&  \Delta \;  \vet{w}_3\cdot \vet{v}(\vet{x}) \;\delta_{\vet{x}+\vet{a}_2,\vet{y}}\delta_{\sigma\sigma'},
\end{eqnarray} 
corresponding to the three nearest-neighbor bonds. The threefold rotation $C_3$ relates them as $\vet{w}_2 = G_2 X \vet{w}_1$ and $\vet{w}_3 = (G_2 X)^2 \vet{w}_1$. The sixfold rotations give the relations $G_2X^T \vet{w}_1 = \pm \vet{w}_3 = \pm   (G_2 X)^2 \vet{w}_1$ and $X \vet{w}_1 = \pm \vet{w}_2 = \pm G_2 X \vet{w}_1$. Both lead to the same constraint $G_3 \vet{w}_1 = \pm \vet{w}_1$. The diagonal reflections all impose the constraint $G_3 Y \vet{w}_1 = \pm \vet{w}_1$, while the vertical reflections impose the constraint $ Y \vet{w}_1 = \pm \vet{w}_1$. All these constraints can be solved to obtain solutions for $\vet{w}_1 $. The equation $G_3 \vet{w}_1 = \vet{w}_1$ gives the solution $\vet{w}_1  = (0,1,0)$, fully specifying honeycomb $F_1$ bond order. 

Instead, $G_3 \vet{w}_1 = -\vet{w}_1$ gives $\vet{w}_1  = (1,0,\pm 1)$. We find that the solution $\vet{w}_1  = i(1,0, 1)$ corresponds to $F_2$ flux order by evaluating the symmetry constraints. Fourier transforming of the real space expressions using $\vet{w}_1  = i(1,0, 1)$ ($\vet{w}_{2,3}$ are obtained by $C_3$) yields the condensate functions of Eq.~\eqref{eq:f2honey}.


\section{Extended point groups~\label{app:gt}}

The purpose of this appendix is to explain the concept of extended point groups in somewhat more detail. In addition, for convenience, we give the character tables of the (extended) point groups used in the main text.

The group of all spatial transformations leaving a given crystal lattice invariant is the space group $S$. Following Appendix~\ref{app:latsym}, $S$ consists of all translations $T$, an Abelian subgroup of $S$, and the point group $G$. The point group can be viewed as the factor group of the space group, i.e., $G= S/T$. The translation subgroup $T$ is generated by the elements $T(\vet{a}_1)$ and $T(\vet{a}_2)$, corresponding to the two elementary lattice vectors $\vet{a}_1$ and $\vet{a}_2$. 

As explained in the main text (see Sec.~\ref{sec:class}), the \emph{extended} point group is the point group of an enlarged unit cell, where the unit cell is chosen so as to support all density wave patterns compatible with a predetermined set of wave vectors. This defines a modified translation subgroup $\widetilde{T}$: the group of all translations that preserve the enlarged unit cell, i.e., map the enlarged unit cell to itself. Another way of saying this, is that the new group of translations is given by all translations $\{ T(\vet{y}) \}$ satisfying $e^{i \vet{y}\cdot \vet{Q}_\mu} =1$, where $\vet{Q}_\mu$ are the specified ordering vectors. The group $\widetilde{T}$ is smaller than $T$. Given the new translation subgroup, a new point group, i.e., the extended point group $\widetilde{G} $, is obtained in the same way as before, by taking the factor group $\widetilde{G} = S/ \widetilde{T}$. The point group $\widetilde{G}$ is larger than $G$, as it contains elements of $T$ no longer part of $\widetilde{T}$. For instance, if $t_1 \equiv T(\vet{a}_1)$ is no longer part of $\widetilde{T}$, it belongs to the extended point group $\widetilde{G}$. 

Here we consider square and hexagonal symmetry, given by the point groups $C_{4v}$ and $C_{6v}$, respectively. We have adopted the convention that extended point groups are denoted as $C'''_{nv}$, where the number of primes indicates the number of translations added to the point group $C_{nv}$. This is equal to the number of inequivalent translations that are part of $T$ but not $\widetilde{T}$. Two translations which are part of $T$ but not $\widetilde{T}$ are equivalent if their their difference is part of $\widetilde{T}$.

Both for the case of the square and the hexagonal Bravais lattices we consider density wave formation at wave vectors $\vet{Q}_\mu$ satisfying $2 \vet{Q}_\mu=0$ and $\vet{Q}_1+\vet{Q}_2+\vet{Q}_3=0$, as explained in Sec.~\ref{sec:class}. This implies translational symmetry breaking such that $\widetilde{T}$ is generated by $T(2\vet{a}_1)$ and $T(2\vet{a}_2)$. As a result, the translations $T(\vet{a}_1) \equiv t_1$, $T(\vet{a}_2) \equiv t_2$ and $T(\vet{a}_1+\vet{a}_2) \equiv t_3$ are added to the point group. To illustrate this, let us take the hexagonal group $C_{6v}$ as an example. It has $12$ elements, and the group $C'''_{6v}$ (three primes indicate three broken elementary translations), which also contains $t_{1,2,3}$, consists of $48$ elements, i.e. $48  = 12 + 3 \times 12$. Algebraic properties of the elements follow from $RT(\vet{x})  = T(R\vet{x})R$ and the fact that $t_it_j = |\epsilon_{ijk}|t_k$. Conjugacy classes and the full character table of $C'''_{6v}$ can be obtained in the standard way. As the point group $C_{6v}$ is a proper subgroup of $C'''_{6v}$ all irreducible representations of $C_{6v}$ will also be irreducible representations of $C'''_{6v}$, in addition to new representations originating from the nontrivial translations. The character tables of the groups $C'''_{6v}$ (hexagonal) and $C'''_{4v}$ (square) are given in Tables~\ref{tab:cppp6v} and~\ref{tab:cppp4v}, respectively. 

A different extended point group is obtained if the hexagonal lattice unit cell is tripled. In that case, the translations $T(\vet{a}_1) \equiv t_1$, $T(\vet{a}_1+\vet{a}_2) \equiv t_2$ (i.e., a redefinition of the $t_i$) are added to the point group, leading to the group $C''_{6v}$. The character table is obtained in the same way and can be found in Ref.~\onlinecite{basko08}.

\subsection{Lattice angular momentum basis functions\label{app:basis}}

\begin{table}[t]
\centering
\begin{ruledtabular}
\begin{tabular}{cccc}
Rep. & Type & Label  & Expression \\ 
\hline
$A_1$  & $s'$ & $\lambda_s(\vet{k})$ & $(\cos k_1+\cos k_2 + \cos k_3)/\sqrt{3}$ \\
$B_1$  & $f$  & $\lambda_f(\vet{k})$  & $(\sin k_1+\sin k_2 + \sin k_3)/\sqrt{3}$   \\
$E_1$  &  $p_x$ & $\lambda_{p_1}(\vet{k})$  & $(\sin k_1+\sin k_2 -2\sin k_3)/\sqrt{6}$  \\
 &  $p_y$  &  $\lambda_{p_2}(\vet{k})$ & $(\sin k_1-\sin k_2)/\sqrt{2} $  \\
$E_2$   & $d_{x^2-y^2}$ &$\lambda_{d_1}(\vet{k})$& $(\cos k_1+\cos k_2 -2\cos k_3 )/\sqrt{6}$  \\
 & $d_{xy}$ & $\lambda_{d_2}(\vet{k})$  & $(\cos k_1-\cos k_2 )/\sqrt{2}$  \\
\end{tabular}
\end{ruledtabular}
 \caption{Lattice angular momentum functions transforming as representations of $C_{6v}$. They apply to the triangular lattice (nearest neighbors) and honeycomb lattice (next-nearest neighbors). We use the definition $k_i = \vet{k}\cdot \vet{a}_i$. Note that $(\lambda_{p_1},\lambda_{p_2}) \sim (k_x,k_y) $ and $(\lambda_{d_1},\lambda_{d_2}) \sim (k_x^2-k_y^2,2k_xk_y)$ when expanded in $\vet{k}$. }
\label{tab:trifunctions}
\end{table}

\begin{table}[t]
\centering
\begin{ruledtabular}
\begin{tabular}{cccc}
Rep. & Type & Label  & Expression \\ 
\hline
$A_1$  & $s'$ & $\lambda_s(\vet{k})$ & $(e^{-i k_1} + e^{-i k_2}+e^{-i k_3})/\sqrt{3}$ \\
$E_2$   & $d_{x^2-y^2}$ &$\lambda_{d_1}(\vet{k})$& $(e^{-i k_1} + e^{-i k_2}-2e^{-i k_3} )/\sqrt{6}$  \\
 & $d_{xy}$ & $\lambda_{d_2}(\vet{k})$  & $(e^{-i k_1} - e^{-i k_2})/\sqrt{2}$  \\
\end{tabular}
\end{ruledtabular}
 \caption{Nearest neighbor angular momentum functions of the honeycomb lattice transforming as representations of $C_{6v}$. Note that these are written in the tight-binding gauge to make hexagonal symmetry transparent. To obtain the gauge adopted in this work, they should be multiplied with the gauge factor $e^{i\vet{\delta}_1\cdot\vet{k}}\equiv e^{i \varphi( \vet{k})}$.}
\label{tab:honeyfunctions}
\end{table}

The lattice angular momentum form factor functions, used in Sec.~\ref{ssec:condtri} to express condensate functions, are given in Table~\ref{tab:trifunctions} and~\ref{tab:honeyfunctions}. Table~\ref{tab:trifunctions} lists the functions that transform as representations of $C_{6v}$ and correspond to a triangular lattice spanned by vectors $\vet{a}_i$. Therefore, they apply to each sublattice of the honeycomb lattice and describe next-nearest neighbor for factors in that case.

Table~\ref{tab:honeyfunctions} lists the lattice angular momentum form factors of the honeycomb lattice. These form factors correspond to nearest neighbor bonds. To make the hexagonal symmetry transparent, they are written in the tight-binding gauge (see Appendix~\ref{app:latsym}). When used in expressions for condensate functions they should be multiplied with the gauge factor $e^{i\vet{\delta}_1\cdot\vet{k}}\equiv e^{i \varphi( \vet{k})}$.

\subsection{Character tables}

For completeness and convenience, here we reproduce the character tables of the extended point groups $C'''_{4v}$ (see Table \ref{tab:cppp4v}) and $C'''_{6v}$ (see Table \ref{tab:cppp6v}).

\begin{table*}[t]
\centering
\begin{ruledtabular}
\begin{tabular}{c|rrrrrrrrrrrrrr}
 Conjugacy class         &   $\mathcal{C}'''_1$   & $\mathcal{C}'''_2$  & $\mathcal{C}'''_3$ & $\mathcal{C}'''_4$  & $\mathcal{C}'''_5$ & $\mathcal{C}'''_6$ & $\mathcal{C}'''_7$  &$\mathcal{C}'''_8$ & $\mathcal{C}'''_9$ & $\mathcal{C}'''_{10}$ & $\mathcal{C}'''_{11}$  &$\mathcal{C}'''_{12}$ & $\mathcal{C}'''_{13}$ & $\mathcal{C}'''_{14}$ \\ 
    Point group  $C'''_{4v} $       &      &  & &     &  &  & &   & &  &&&& \\  [1ex]
\hline 
 $A_1$ & $1$& $1$ & $1$ & $1$  & $1$ & $1$  &$1$   &$1$ & $1$ &  $1$ &$1$   &$1$ & $1$ &  $1$\\ 
 $A_2$ & $1$& $1$ & $1$ & $1$  & $1$ & $1$  &$1$ & $1$ & $-1$ & $-1$ &$-1$   &$-1$ & $-1$ &  $-1$\\ 
 $B_1$ & $1$& $1$ & $1$ & $1$  & $1$ & $1$  &$-1$ & $-1$ & $1$ & $1$  &$1$   &$1$ & $-1$ &  $-1$\\
 $B_2$ & $1$& $1$ & $1$ & $1$  & $1$ & $1$  &$-1$ & $-1$ & $-1$ & $-1$ &$-1$   &$-1$ & $1$ &  $1$\\
 $E_1$ & $2$& $2$ & $2$& $-2$ & $-2$& $-2$  &$ 0$  & $0$ & $0$ &  $0$ &$0$   &$0$ & $0$ &  $0$ \\
\hline
 $A'_1$ & $1$& $-1$ & $1$ & $1$ & $-1$ & $1$  &$1$   &$-1$ & $1$ & $-1$ &$-1$   &$1$ & $1$ &  $-1$\\ 
 $A'_2$ & $1$& $-1$ & $1$ & $1$ & $-1$ & $1$  &$1$ & $-1$ & $-1$ & $1$ &$1$   &$-1$ & $-1$ &  $1$\\ 
 $B'_1$ & $1$& $-1$ & $1$ & $1$ & $-1$ & $1$  &$-1$  & $1$ & $1$ & $-1$  &$-1$   &$1$ & $-1$ &  $1$\\
 $B'_2$ & $1$& $-1$ & $1$ & $1$ & $-1$ & $1$  &$-1$ & $1$ & $-1$ & $1$ &$1$   &$-1$ & $1$ &  $-1$\\

 $E'_1$ & $2$& $-2$ & $2$  & $-2$ & $2$& $-2$  &$ 0$  & $0$ & $0$ &  $0$  &$0$ &$0$ & $0$ &  $0$\\
 $E_2$  & $2$& $0$  & $-2$ & $2$  & $0$& $-2$  &$ 0$  & $0$ & $-2$ &  $0$  &$0$ &$2$ & $0$ &  $0$\\
 $E_3$  & $2$& $0$  & $-2$ & $2$  & $0$& $-2$  &$ 0$  & $0$ & $2$ &  $0$  &$0$ &$-2$ & $0$ &  $0$\\
 $E_4$  & $2$& $0$  & $-2$ & $-2$ & $0$& $2$   &$ 0$  & $0$ & $0$ &  $2$  &$-2$ &$0$ & $0$ &  $0$\\
 $E_5$  & $2$& $0$  & $-2$ & $-2$ & $0$ & $2$  &$ 0$  & $0$ & $0$ &  $-2$  &$2$ &$0$ & $0$ &  $0$
\end{tabular}
\end{ruledtabular}
 \caption{Character table of the point group $C'''_{4v}$. Translations $t_1$ and $t_2$ correspond to $T(\vet{a}_1)$ and $T(\vet{a}_2)$, respectively, and $t_3 = T(\vet{a}_1+\vet{a}_2)$. The conjugacy classes consist of the elements: $\mathcal{C}'''_1 = \{ I \}$, $\mathcal{C}'''_2 = \{ t_1, t_2 \}$, $\mathcal{C}'''_3 = \{ t_3 \}$, $\mathcal{C}'''_4 = \{ C_2 \}$, $\mathcal{C}'''_5 = \{ t_1C_2, t_2C_2 \}$, $\mathcal{C}'''_6 = \{  t_3C_2 \}$, $\mathcal{C}'''_7 = \{ C_4, C^{-1}_4 , t_3C_4, t_3C^{-1}_4\}$, $\mathcal{C}'''_8 = \{ t_1C_4, t_1C^{-1}_4 , t_2C_4, t_2C^{-1}_4  \}$, $\mathcal{C}'''_9 = \{ \sigma_{v1}, \sigma_{v2} \}$,  $\mathcal{C}'''_{10} = \{ t_1\sigma_{v1}, t_2\sigma_{v2} \}$, $\mathcal{C}'''_{11} = \{  t_2\sigma_{v1}, t_1\sigma_{v2}  \}$, $\mathcal{C}'''_{12} = \{ t_3\sigma_{v1}, t_3\sigma_{v2} \}$, $\mathcal{C}'''_{13} = \{  \sigma_{d1},\sigma_{d2},  t_3\sigma_{d1}, t_3\sigma_{d2} \}$ and $\mathcal{C}'''_{14} = \{ t_1\sigma_{d1},t_1\sigma_{d2},  t_2\sigma_{d1}, t_2\sigma_{d2}  \}$. The character table is taken from Ref.~\onlinecite{serbyn13}. Notation is altered with respect to Ref.~\onlinecite{serbyn13} to be consistent with the notation and definitions of this work.}
\label{tab:cppp4v}
\end{table*}

\begin{table*}[t]
\centering
\begin{ruledtabular}
\begin{tabular}{c|rrrrrrrrrr}
 Conjugacy class         &   \multicolumn{1}{c}{ $\mathcal{C}'''_1$ }   &  \multicolumn{1}{c}{ $\mathcal{C}'''_2$ }   &  \multicolumn{1}{c}{ $\mathcal{C}'''_3$ } & \multicolumn{1}{c}{ $\mathcal{C}'''_4$ }   & \multicolumn{1}{c}{ $\mathcal{C}'''_5$ } & \multicolumn{1}{c}{ $\mathcal{C}'''_6$ }  & \multicolumn{1}{c}{ $\mathcal{C}'''_7$ }  &\multicolumn{1}{c}{ $\mathcal{C}'''_8$ }  & \multicolumn{1}{c}{ $\mathcal{C}'''_9$ }  & \multicolumn{1}{c}{ $\mathcal{C}'''_{10}$ }  \\ 
\cline{2-11}
      Point group         &      &  $t_1$, $t_2$ & &  $t_1C_2$, $t_2 C_2$   & $t_i C_3$, $t_iC^{-1}_3$ & $t_i C_6$, $t_iC^{-1}_6$ & $ 3\sigma_v$, $t_1\sigma_{v2}$ & $ t_1\sigma_{v}$, $ t_2\sigma_{v}$  &$3\sigma_{d}$, $t_2\sigma_{d1}$ &  $t_1\sigma_{d1}$, $t_3\sigma_{d1}$  \\
  $C'''_{6v} $   & $ I $& $t_3$ & $C_2$  &$t_3 C_2$  & $C_3$, $C^{-1}_3$ & $C_6$, $C^{-1}_6$  & 
 $t_2 \sigma_{v3} $,  $t_3 \sigma_{v1}$ &  $t_2 \sigma_{v2} $,  $t_3 \sigma_{v2}$ &   $t_3 \sigma_{d2}$,  $t_1\sigma_{v3}$ &   $t_1\sigma_{d2}$, $t_2\sigma_{d2}$   \\ 
&  &  &   &  &  &  & 
 &  $t_1 \sigma_{v3} $,  $t_3 \sigma_{v3}$ &    &   $t_2\sigma_{d3}$, $t_3\sigma_{d3}$   \\ [1ex]
\hline 
 $A_1$ & $1$& $1$ & $1$ & $1$  & $1$ & $1$  &$1$   &$1$ & $1$ &  $1$ \\ 
 $A_2$ & $1$& $1$ & $1$ & $1$  & $1$ & $1$  &$- 1$ & $-1$ & $-1$ & $-1$ \\ 
 $B_1$ & $1$& $1$ & $-1$& $-1$ & $1$ & $-1$ &$ 1$  & $1$ & $-1$ & $-1$  \\
 $B_2$ & $1$& $1$ & $-1$& $-1$ & $1$ & $-1$ &$- 1$ & $-1$ & $1$ & $1$ \\
 $E_1$ & $2$& $2$ & $-2$& $-2$ & $-1$& $1$  &$ 0$  & $0$ & $0$ &  $0$ \\
 $E_2$ & $2$& $2$ & $2$ & $2$  & $-1$& $-1$ &$ 0$  & $0$ & $0$ &  $0$ \\
\hline
 $F_1$ & $3$& $-1$& $3$ & $-1$ & $0$& $0$  &$ 1$  & $-1$ & $1$ & $-1$  \\
 $F_2$ & $3$& $-1$& $3$ & $-1$ & $0$& $0$  &$ -1$ & $1$ & $-1$ &  $ 1$ \\
 $F_3$ & $3$& $-1$& $-3$& $1$  & $0$ & $0$  &$ 1$  & $-1$ & $-1$ &  $1$  \\
 $F_4$ & $3$& $-1$& $-3$& $1$  & $0$ & $0$  &$ -1$ & $1$ & $1$ & $-1$
\end{tabular}
\end{ruledtabular}
 \caption{The point group $C'''_{6v}$. Translations $t_1$ and $t_2$ correspond to $T(\vet{a}_1)$ and $T(\vet{a}_2)$, respectively. $t_3 = T(\vet{a}_1+\vet{a}_2)$. The irreducible representations that arise as a consequence of the added translations are $F_1$, $F_2$, $F_3$ and $F_4$, all three-dimensional.}
\label{tab:cppp6v}
\end{table*}

\end{document}